\documentclass{aa}  

\usepackage{hyperref}

\usepackage{xcolor}
\definecolor{dark-red}{rgb}{0.9,0.0,0.0}
\definecolor{dark-blue}{rgb}{0.15,0.15,0.9}
\definecolor{dark-green}{rgb}{0.15,0.8,0.15}
\definecolor{medium-blue}{rgb}{0,0,0.9}
\hypersetup{
    colorlinks, linkcolor=red,
    citecolor={dark-blue} , urlcolor={medium-blue}
}

\usepackage{graphicx}
\usepackage{txfonts}
\begin{document}
   \title{Precise radial velocities of giant stars \\
   }
\subtitle{VI. A possible 2:1 resonant planet pair around the K giant star $\eta$~Cet 
      \thanks{Based on observations collected at Lick Observatory, University of
California.} \fnmsep \thanks{Based on observations collected at the European Southern Observatory, 
Chile, under program IDs 088.D-0132, 089.D-0186, 090.D-0155 and 091.D-0365.}}
 
    \author{Trifon Trifonov\inst{1}, Sabine Reffert\inst{1}, Xianyu Tan\inst{2}$^{,}$\inst{3}, 
    Man Hoi Lee\inst{2}$^{,}$\inst{4}, 
    \and Andreas Quirrenbach\inst{1}}
 
   \institute{ZAH-Landessternwarte,\ K\"{o}nigstuhl 12, 69117 Heidelberg, Germany\\ \vspace{-3mm}
              \and
Department of Earth Sciences, The University of Hong Kong, Pokfulam Road, Hong Kong\\ \vspace{-3mm}
              \and
Department of Planetary Sciences and Lunar and Planetary Laboratory,
The University of Arizona, 1629 University Blvd., Tucson, AZ 85721 USA\\ \vspace{-3mm}
              \and
Department of Physics, The University of Hong Kong, Pokfulman Road, Hong Kong\\ 
}
   
   \date{Received October 22, 2013; accepted June 20, 2014}


  \abstract
   {
   We report the discovery of a new planetary system around the K giant $\eta$~Cet (HIP 5364, HD 6805, HR 334) based on 
   118 high-precision optical radial velocities taken at Lick Observatory since July 2000. Since October 2011 an
   additional nine near-infrared Doppler measurements have been taken using the ESO CRIRES spectrograph (VLT, UT1).  
    The visible data set shows two clear periodicities.    
    Although we cannot completely rule out that the shorter period is due to rotational modulation
    of stellar features, the infrared data show the same variations as in the optical, which strongly supports that the 
    variations are caused by two planets.  
   Assuming the mass of  $\eta$~Cet to be $1.7~$$M_\odot$, the best edge-on coplanar dynamical fit to the data 
   is consistent with two massive planets ($m_b~\sin i~=~2.6~\pm~0.2~$$M_{\mathrm{Jup}}$,
   $m_c~\sin i~=~3.3~\pm~0.2~$$M_{\mathrm{Jup}}$), with periods of $P_b  = 407 \pm3$~days and $P_c = 740 \pm5$~days
   and eccentricities of $e_b = 0.12\pm0.05$ and $e_c = 0.08\pm0.03$. 
   These mass and period ratios suggest possible strong interactions between the planets,
   and a dynamical test is mandatory.
   We tested a wide variety of edge-on coplanar and inclined planetary configurations for stability, which
   agree with the derived radial velocities.
   We find that for a coplanar configuration there are several isolated stable solutions
   and two well-defined stability regions.
   In certain orbital configurations with moderate $e_b$ eccentricity,
   the planets can be effectively trapped in an anti-aligned 2:1 mean motion resonance that stabilizes the system.
   A much larger non-resonant stable region exists in low-eccentricity parameter space,
   although it appears to be much farther from the best fit than the 2:1 resonant region.
   In all other cases, the system is categorized as unstable or chaotic.
   Another conclusion from the coplanar inclined dynamical test is that the planets can be at most a factor of
   $\sim$ 1.4 more massive than their suggested minimum masses. 
   Assuming yet higher inclinations, and thus larger planetary masses, 
   leads to instability in all cases. 
   This stability constraint on the inclination excludes the possibility of two brown dwarfs, and
   strongly favors a planetary system.
   }
   \titlerunning{Precise Radial Velocities of Giant Stars} 
   \authorrunning{Trifon Trifonov et al.}
   \maketitle

\section{Introduction}
Until May 2014, 387 known multiple planet systems were reported in the literature (www.exoplanets.org),
and their number is constantly growing.
The first strong evidence for a multiple planetary system around a main-sequence star was reported by 
\citet{Butler3}, showing that together with the 4.6 day-period radial velocity signal of $\upsilon$ And 
\citep{Butler2}, two more long-period, substellar companions can be derived from the Doppler curve.
Later, a second Jupiter-mass planet was found to orbit the star~47~UMa \citep{Fischer},~and another 
one the G star HD12661 \citep{Fischer3}. 

Interesting cases also include planets locked in mean motion resonance (MMR), 
such as the short-period 2:1 resonance pair around~GJ~876 \citep{Marcy}, 
 the 3:1 MMR planetary system around HD~60532 \citep{Desort,Laskar2009}, or the
3:2 MMR system around HD~45364 \citep{Correia}. 
Follow-up radial velocity observations of well-known planetary pairs showed evidence that
some of them are actually part of higher-order multiple planetary systems. 
For example, two additional long-period planets are orbiting around GJ 876 \citep{Rivera},
and up to five planets are known to orbit 55~Cnc \citep{Fischer2}.
More recently, \citet{Lovis} announced a very dense, but
still well-separated low-mass planetary system around the solar-type star HD~10180.
\citet{Tuomi} claimed that there might even be nine planets in this system, which would make it the most compact and 
populated extrasolar multiple system known to date.

Now, almost two decades since the announcement of 51~Peg~b \citep{Mayor}, 
55 multiple planetary systems have been found using high-precision Doppler spectroscopy (www.exoplanets.org).
Another 328 multiple planetary systems have been found with the transiting technique,
the vast majority of them with the Kepler satellite.
Techniques such as direct imaging  \citep{Marois} and micro-lensing
\citep{Gaudi} have also proven to be successful in detecting multiple extrasolar planetary systems. 

The different techniques for detecting extrasolar planets and the combination between them shows 
that planetary systems appear to be very frequent in all kinds of stable configurations. 
Planetary systems are found to orbit around stars with different ages 
and spectral classes, including binaries \citep{Lee} and even pulsars \citep{Wolszczan}.
Nevertheless, not many multiple planetary systems have been found around evolved giant stars so far.
The multiple planetary systems appear to be a very small fraction of the planet
occurrence statistics around evolved giants, which
are dominated by single planetary systems.
Up to date there is only one multiple planetary system candidate known around an evolved star 
\citep[HD~102272,][]{Niedzielski_a}, 
and two multiple systems consistent with brown dwarf mass companions around BD~+20~2457 \citep{Niedzielski_b}
and $\nu$ Oph \citep{Quirrenbach,Sato}.

In this paper we present evidence for two Jovian planets orbiting the K giant $\eta$~Cet based on 
precise radial velocities. We also carry out an extensive stability analysis to demonstrate that the
system is stable and to further constrain its parameters.

The outline of the paper is as follows: in~\S\ref{Observations and stellar characteristics}
we introduce the stellar parameters for $\eta$~Cet and
describe our observations taken at Lick observatory and at the VLT. 
Section~\S\ref{Orbital fit} describes the derivation
of the spectroscopic orbital parameters. In~\S\ref{Stability tests} 
we explain our dynamical stability calculations, and in \S\ref{Discussion eta cet} we
discuss the possible origin of the $\eta$~Cet system and the population of planets around giants.
Finally, we provide a summary in Section~\S\ref{Summary}.

\section{Observations and stellar characteristics}
\label{Observations and stellar characteristics}

\subsection{K giant star $\eta$ Cet}

 $\eta$~Cet (= HIP 5364,  HD 6805, HR 334) is a bright ($V$ = 3.46 mag), red giant clump star 
($B~-~V$~=~1.16). It is located at a distance of 37.99 $\pm$ 0.20 pc \citep{Leeuwen} and
flagged in the Hipparcos catalog as photometrically constant.

\citet{Luck} proposed T$_{\mathrm{eff}}$ = 4425 $\pm$ 100 K, derived from photometry, and
$\log g$~=~2.65 $\pm$ 0.25 $[\mathrm{cm\cdot s}^{-2}]$ estimated from the ionization balance
between Fe I and Fe II lines in the spectra.
 \citet{Luck} derived [Fe/H]~=~0.16~$\pm$~0.05, and a mass  of $M$~=~1.3~$\pm$~0.2~$M_\odot$.
The more recent study of  $\eta$~Cet by \citet{Berio} derived the 
stellar parameters as T$_{\mathrm{eff}}$~=~4356~$\pm$~55~K, luminosity
$L$~=~74.0~$\pm$~3.7~$L_\odot$, and the estimated radius as $R$~=~15.10~$\pm$~0.10~$R_\odot$.
\citet{Berio} roughly estimate the mass of  $\eta$~Cet to be $M$ = 1.0$-$1.4~$M_\odot$ by comparing
its position in the Hertzsprung–Russell (HR) diagram with evolutionary tracks of solar metallicity.

By using a Lick template spectrum without iodine absorption cell lines, 
\citet{Hekker2} estimated the metallicity of $\eta$~Cet to be [Fe/H] = 0.07 $\pm$ 0.1.
Based on this metallicity and the observed position in the HR diagram, a trilinear interpolation in the 
evolutionary tracks and isochrones \citep{Girardi2000} yields $T_{\mathrm{eff}}$~=~4529~$\pm$~19~K, 
$\log g$~=~2.36~$\pm$~0.05~$[\mathrm{cm\cdot s}^{-2}]$, 
$L$~=~77.1~$\pm$~1.1~$L_{\odot}$ and $R$~=~14.3~$\pm$~0.2~$R_{\odot}$ \citep{Reffert2014}.


\begin{table}[htp]

\caption{Stellar properties of $\eta$~Cet.} 
\label{table:phys_param}    

\resizebox{0.34\textheight}{!}{\begin{minipage}{\textwidth/2} 

\vspace{-0.2cm}
\centering          
\begin{tabular}{ l l r}     
\hline\hline  \noalign{\vskip 0.5mm}        
  Parameter   &$\eta$~Cet   &  reference \\  
\hline    \noalign{\vskip 0.5mm}                   
    Spectral type                           & K1III             & \citet{Gray} \\ 
                                            & K2III CNO.5       & \citet{Luck} \\     
   $m_v$   [mag]                            & 3.46              & \citet{Leeuwen} \\
   $B - V$                                  & 1.161 $\pm$ 0.005 & \citet{Leeuwen} \\
   Distance  [pc]                           & 37.99   $\pm$ 0.20   & \citet{Leeuwen} \\                                        
   $\pi$  [mas]                             & 26.32  $\pm$ 0.14 & \citet{Leeuwen} \\ 
   Mass    [$M_{\odot}$]                    & 1.7  $\pm$ 0.1   & \citet{Reffert2014} \\
                                            & 1.3  $\pm$ 0.2    & \citet{Luck} \\   
                                            & 1.2  $\pm$ 0.2    & \citet{Berio} \\ 
   Luminosity    [$L{_\odot}$]              & 77.1  $\pm$ 1.1  & \citet{Reffert2014} \\
                                            & 74.0   $\pm$ 3.7  & \citet{Berio} \\
   Radius    [$R_{\odot}$]                  & 14.3   $\pm$ 0.2  & \citet{Reffert2014}  \\
                                            & 15.1   $\pm$ 0.1  & \citet{Berio} \\  
   $T_{\mathrm{eff}}$~[K]                    & 4528 $\pm$ 19     & \citet{Reffert2014} \\
                                            & 4563 $\pm$ 82     &\citet{Prugniel} \\
                                            & 4425 $\pm$ 100    & \citet{Luck}   \\ 
                                            & 4356 $\pm$ 55     & \citet{Berio}   \\
   $\log g~[\mathrm{cm\cdot s}^{-2}]$       & 2.36 $\pm$ 0.05   & \citet{Reffert2014} \\
                                            & 2.61 $\pm$ 0.21   &\citet{Prugniel} \\
                                            & 2.65 $\pm$ 0.25   & \citet{Luck}   \\ 
   {}[Fe/H]                                 & 0.07 $\pm$ 0.1    & \citet{Hekker2}  \\
                                            & 0.12 $\pm$ 0.08   &\citet{Prugniel} \\    
                                            & 0.16 $\pm$ 0.05   & \citet{Luck}   \\ 
   $v~\sin i$~$[$km\,s$^{-1}]$              & 3.8  $\pm$ 0.6$^\alpha$    & \citet{Hekker2}   \\                                          
   RV$_{absolute}$   $[$km\,s$^{-1}]$       & 11.849            & this paper     \\ 
   
\hline\hline \noalign{\vskip 0.5mm}   

\end{tabular}

    \makebox[0.999\textwidth][l]{\tiny $\alpha$ - we estimated this using $\sigma_{FWHM}$ and $\sigma_{v_{mac}}$
    given in \citet{Hekker2}} \par      
\vspace{-0.3cm}
\end{minipage}}

\end{table}
 
We determined  the probability of $\eta$~Cet to be on the red giant branch (RGB) 
or on the horizontal branch (HB) by generating 10\,000 positions with (B -- V, $M_V$, [Fe/H]) 
consistent with the error bars on these quantities, and
derived the stellar parameters via a comparison with interpolated evolutionary tracks. 
Our method for deriving stellar parameters for all G and K giant stars monitored at Lick Observatory,
including $\eta$~Cet, is described in more detail in \citet{Reffert2014}.

Our results show that $\eta$~Cet has a 70\% probability to be on the RGB with a resulting mass of
$M$ = 1.7 $\pm$ 0.1 $M_\odot$. If $\eta$~Cet were on the HB, the mass would be 
$M$ = 1.6 $\pm$ 0.2 $M_\odot$.
Here we simply use the mass with the highest probability.
All stellar parameters
are summarized in Table \ref{table:phys_param}.

\subsection{Lick data set}
Doppler measurements for $\eta$~Cet have been obtained since July 2000 as part of our precise (5 -- 8~m\,s$^{-1}$) 
Doppler survey of 373 very bright (V $\leq$ 6 mag) G and K giants. 
The program started in June 1999 using the 0.6~m  
Coud\'{e} Auxiliary Telescope (CAT) with the Hamilton \'{E}chelle Spectrograph at Lick Observatory.  
The original goal of the program was to study the intrinsic radial velocity variability in K giants, and 
to demonstrate that the low levels of stellar jitter make these stars a good choice for astrometric reference objects for 
the Space Interferometry Mission (SIM) \citep{Frink}.
However, the low amplitude of the intrinsic jitter of the selected K giants, together with the precise and regular
observations, makes this survey sensitive to variations in the radial velocity that might be caused by extrasolar planets.

All observations at Lick Observatory have been taken with the iodine cell placed in the light path at the entrance of the spectrograph.
This technique provides us with many narrow and very well defined iodine spectral lines, which are used as references,
and it is known to yield precise Doppler shifts down to 3 m\,s$^{-1}$ or even better for dwarf stars \citep{Butler}.
The iodine method is not discussed in this paper; instead we refer to \citet{Marcy2},\citet{Valenti}, 
and \citet{Butler}, where more details about the technique and the data reduction can be found.   

\begin{figure}[htp]
\resizebox{\hsize}{!}{\includegraphics{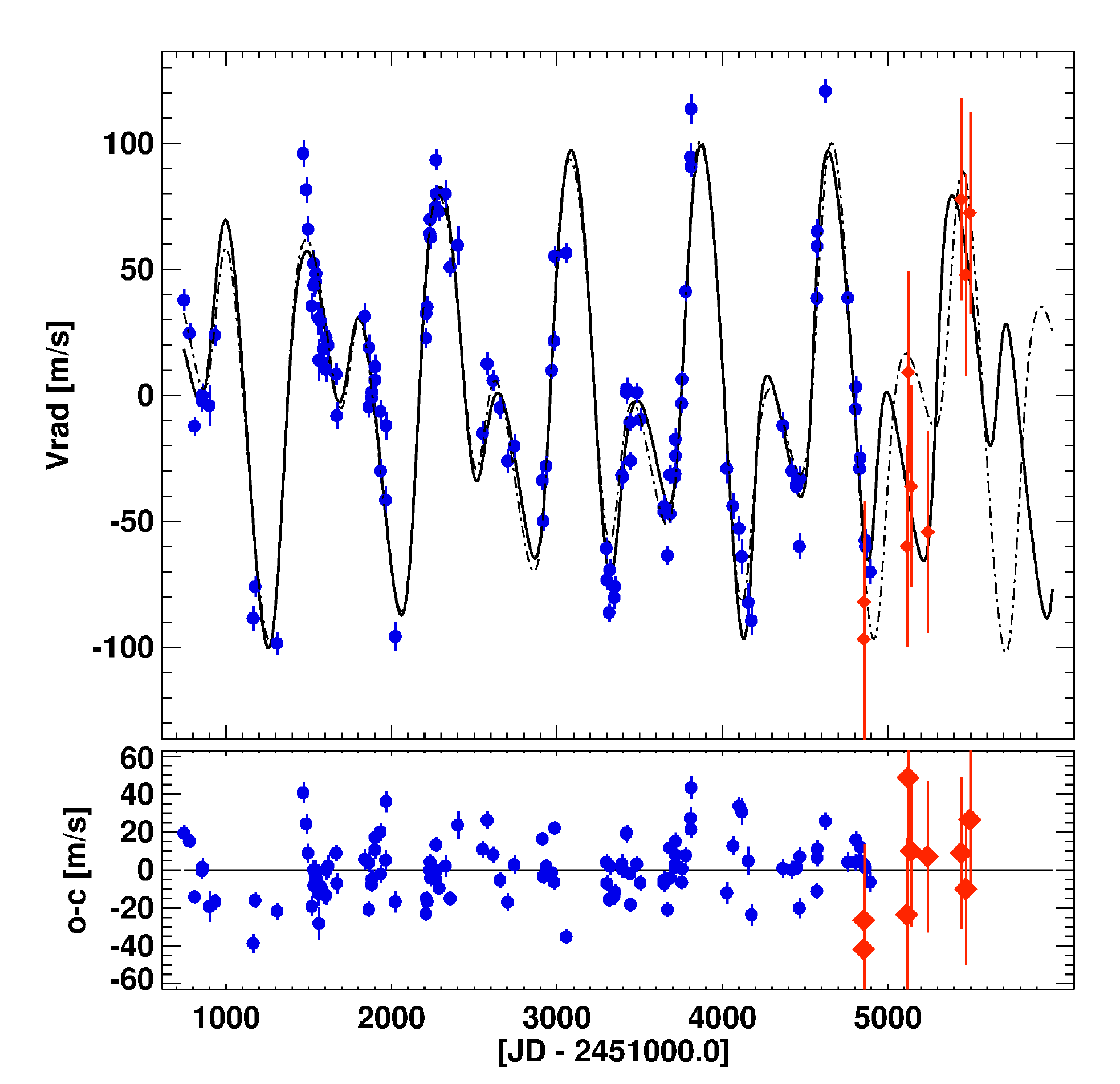}}

\caption{$Top~panel$: radial velocities measured at Lick Observatory ($blue~circles$), along with 
error bars, covering about
11 years from July 2000 to October 2011. Two best fits to the Lick data are overplotted:
a double-Keplerian fit ($dot~dashed$) and
the best dynamical edge-on coplanar fit ($solid~line$). The two fits are not consistent in later
epochs because of the gravitational interactions considered in the dynamical model.
Despite the large estimated errors, the data from CRIRES ($red~diamonds$) seem to follow the best~fit prediction
from the dynamical fit.
$Bottom~panel$: no systematics are visible in the residuals. 
The remaining radial velocity scatter has a standard deviation of 15.9 m\,s$^{-1}$, most likely caused by 
rapid solar--like $\it{p}$ -- mode oscillations.}
\label{FigGam:rv}
\centering

\end{figure}


The wavelength coverage of the Hamilton spectra extends from 3755 to 9590 \AA, with 
a resolution of $R$ $\approx$ 60\,000 
in the wavelength range from 5000 to 5800 \AA, where most of the iodine lines can be found and where the radial velocities are measured.
The typical exposure time with the 0.6\,m CAT is 450 seconds, which results in a signal-to-noise ratio of about 100, 
reaching a radial velocity precision of better than 5 m\,s$^{-1}$.
The individual radial velocities are listed in Table~\ref{table:rvlick},
together with Julian dates and their formal errors.

\begin{table*}{}

\caption{Measured velocities for $\eta$~Cet and the derived errors }   
\label{table:rvlick} 
\resizebox{0.65\textheight}{!}{\begin{minipage}{\textwidth} 
 
 \begin{tabular}{ crc  p{0.01mm} crc p{0.01mm} crc }
\cline{1-3}\cline{1-3}\cline{5-7}\cline{5-7}\cline{9-11}\cline{9-11}
\noalign{\vskip 0.5mm}
\cline{1-3}\cline{1-3}\cline{5-7}\cline{5-7}\cline{9-11}\cline{9-11}
\noalign{\vskip 0.5mm}


    JD & RV [m\,s$^{-1}$] & $\sigma_{RV}$ [m\,s$^{-1}$] & &JD & RV [m\,s$^{-1}$] & $\sigma_{RV}$  [m\,s$^{-1}$] && JD & RV [m\,s$^{-1}$] & $\sigma_{RV}$  [m\,s$^{-1}$]\\  
\cline{1-3}\cline{5-7}\cline{9-11}
\noalign{\vskip 0.5mm}
2451745.994   &  37.8    & 4.3 &&   2453231.961   &  64.3    & 4.0 &&  2454711.893   & $-$32.4    & 4.1  \\
2451778.895   &  24.7    & 3.7 &&   2453233.909   &  69.9    & 3.9 &&  2454713.873   & $-$31.0    & 4.2  \\
2451808.867   & $-$12.2  & 3.6 &&   2453235.960   &  62.6    & 4.2 &&  2454715.882   & $-$17.4    & 4.6  \\
2451853.766   &  $-$2.1  & 4.1 &&   2453265.899   &  74.7    & 3.8 &&  2454716.875   & $-$24.0    & 4.1  \\
2451856.801   &   0.1    & 4.6 &&   2453268.919   &  93.4    & 4.0 &&  2454753.885   &  $-$3.2    & 3.6  \\
2451898.637   &  $-$4.0  & 8.0 &&   2453269.810   &  80.0    & 3.5 &&  2454755.850   &   6.4      & 3.8  \\
2451932.609   &  23.9    & 3.9 &&   2453286.828   &  73.1    & 3.8 &&  2454777.733   &  41.3      & 4.1  \\
2452163.861   & $-$89.3  & 4.9 &&   2453324.713   &  80.0    & 5.5 &&  2454806.796   &  94.7      & 5.5  \\
2452175.851   & $-$75.7  & 4.0 &&   2453354.654   &  50.9    & 3.8 &&  2454808.770   &  90.8      & 4.3  \\
2452307.598   & $-$98.1  & 4.5 &&   2453400.642   &  59.6    & 7.5 &&  2454809.763   & 113.7      & 6.1  \\
2452465.989   &  96.1    & 5.3 &&   2453550.983   & $-$14.9  & 4.4 &&  2455026.973   & $-$30.0   & 5.7  \\
2452483.985   &  81.6    & 5.0 &&   2453578.964   &  12.6    & 4.6 &&  2455063.965   & $-$43.9    & 5.0  \\
2452494.987   &  66.0    & 5.0 &&   2453613.923   &   6.0    & 4.2 &&  2455099.996   & $-$52.8    & 4.9  \\
2452517.936   &  35.5    & 5.2 &&   2453654.771   &  $-$4.9  & 4.1 &&  2455116.856   & $-$63.9    & 6.8  \\
2452528.910   &  52.4    & 5.2 &&   2453701.724   & $-$26.0  & 4.6 &&  2455154.739   & $-$82.1    & 7.6  \\
2452530.910   &  43.7    & 4.5 &&   2453740.745   & $-$20.1  & 4.7 &&  2455174.788   & $-$89.3    & 5.8  \\
2452541.904   &  45.1    & 4.5 &&   2453911.993   & $-$33.7  & 3.6 &&  2455364.972   & $-$11.9    & 5.0  \\
2452543.883   &  48.3    & 5.0 &&   2453917.013   & $-$49.9  & 3.9 &&  2455419.012   & $-$29.9    & 4.7  \\
2452559.914   &  30.5    & 6.4 &&   2453934.974   & $-$28.0  & 3.9 &&  2455446.847   & $-$35.4    & 3.6  \\
2452561.795   &  14.0    & 8.4 &&   2453968.925   &  9.9    & 3.2 &&  2455446.856   & $-$36.2    & 3.1  \\
2452571.791   &  29.7    & 5.0 &&   2453981.851   &  21.6    & 3.8 &&  2455463.833   & $-$59.8    & 5.2  \\
2452589.810   &  18.3    & 5.7 &&   2453985.899   &  55.2    & 4.0 &&  2455466.842   & $-$33.0    & 4.8  \\
2452603.782   &  10.4    & 4.9 &&   2454055.755   &  56.8    & 3.9 &&  2455569.653   &  40.0      & 4.1  \\
2452605.724   &  22.9    & 4.8 &&   2454297.975   & $-$60.6  & 4.1 &&  2455571.713   &  59.6      & 4.6  \\
2452616.772   &  19.9    & 5.9 &&   2454300.974   & $-$73.2  & 4.0 &&  2455572.671   &  65.6      & 4.8  \\
2452665.604   &   8.5    & 4.2 &&   2454314.939   & $-$86.2  & 3.6 &&  2455621.621   & 121.2      & 4.6  \\
2452668.611   &  $-$8.0  & 5.3 &&   2454318.981   & $-$69.2  & 4.2 &&  2455757.974   &  39.1      & 5.0  \\
2452837.972   &  31.3    & 5.4 &&   2454344.859   & $-$80.2  & 4.3 &&  2455803.919   &  $-$5.4    & 4.3  \\
2452861.963   &  $-$4.6  & 4.1 &&   2454348.917   & $-$75.8  & 4.0 &&  2455806.936   &   3.4      & 4.3  \\
2452863.985   &  19.0    & 5.2 &&   2454391.887   & $-$31.6  & 5.4 &&  2455828.781   & $-$30.0    & 4.2  \\
2452879.918   &  $-$0.9  & 4.2 &&   2454394.775   & $-$32.4  & 4.2 &&  2455831.849   & $-$24.8    & 5.1  \\
2452880.947   &   1.3    & 5.0 &&   2454418.777   &   1.2    & 4.5 && $\star$2455853.844   & $-$96.7 & 40.0 \\
2452898.911   &   6.1    & 4.7 &&   2454420.763   &   2.7    & 4.2 && $\star$2455854.616   &  $-$82.0 & 40.0 \\
2452900.928   &  11.4    & 4.8 &&   2454440.653   & $-$10.6  & 3.4 &&  2455861.842   & $-$57.5  & 4.2  \\
2452932.839   &  $-$6.3  & 4.3 &&   2454443.651   & $-$26.0  & 3.6 &&  2455864.796   & $-$59.9  & 4.4  \\
2452934.804   & $-$29.9  & 4.6 &&   2454481.638   &   1.2    & 3.9 &&  2455892.745   & $-$69.9  & 4.9  \\
2452963.753   & $-$41.5  & 5.2 &&   2454503.651   &  $-$9.5  & 4.3 && $\star$2456113.863   &  $-$60.0   & 40.0 \\ 
2452965.809   & $-$11.9  & 5.4 &&   2454645.984   & $-$44.1  & 4.5 && $\star$2456121.811   &  9.1  & 40.0 \\
2453022.630   & $-$95.6  & 5.6 &&   2454646.968   & $-$46.2  & 4.2 && $\star$2456139.747   &  $-$36.2   & 40.0 \\
2453208.010   &  22.7    & 3.8 &&   2454668.009   & $-$63.5  & 3.7 && $\star$2456239.563   &  $-$54.2   & 40.0 \\
2453209.978   &  32.5    & 3.7 &&   2454681.904   & $-$31.4  & 3.8 && $\star$2456411.910   & 77.7  & 40.0  \\
2453214.961   &  35.2    & 4.2 &&   2454683.970   & $-$47.0  & 3.8 && $\star$2456469.917   & 47.9  & 40.0 \\
              &          &     &&                 &          &     && $\star$2456494.892   & 72.4  & 40.0  \\
\cline{1-3}\cline{1-3}\cline{5-7}\cline{5-7}\cline{9-11}\cline{9-11}
\noalign{\vskip 0.5mm}
\cline{1-3}\cline{1-3}\cline{5-7}\cline{5-7}\cline{9-11}\cline{9-11}
\noalign{\vskip 2.5mm}

  $\star$ - CRIRES data \\
 \makebox[0.1\textwidth][l]{absolute RV$_{CRIRES}$ = 11.849 km\,s$^{-1}$ } \par \\

\end{tabular}
\end{minipage}}
\end{table*}

\subsection{CRIRES data set}

Nine additional Doppler measurements for $\eta$~Cet were taken between October 2011 and July 2013
with the pre-dispersed CRyogenic InfraRed Echelle Spectrograph (CRIRES) mounted at VLT UT1 (Antu), \citep{Kaeufl}.
CRIRES has a resolving power of $\it{R} \approx$~100\,000 
when used with a 0.2\textquotedblright~slit, covering a narrow wavelength region in the 
J, H, K, L or M infrared bands (960 -- 5200 nm).
Several studies have demonstrated that radial velocity measurements  
 with precision between 10 and 35~m\,s$^{-1}$ are possible with CRIRES.
\citet{Seifahrt} reached a precision of $\approx$~35~m\,s$^{-1}$ when using reference spectra of a 
N$_2$O gas cell, and \citet{Bean2} even reached $\approx$ 10 m\,s$^{-1}$ with an ammonia (NH$_3$)
gas-cell. \citet{Huelamo} and \citet{Figueira} showed that the achieved Doppler precision can 
be better than $\approx$~25~m\,s$^{-1}$ when using telluric absorption lines in the $H$ band as reference spectra. 

\begin{figure}[htp]

\centering
\resizebox{\hsize}{!}{\includegraphics{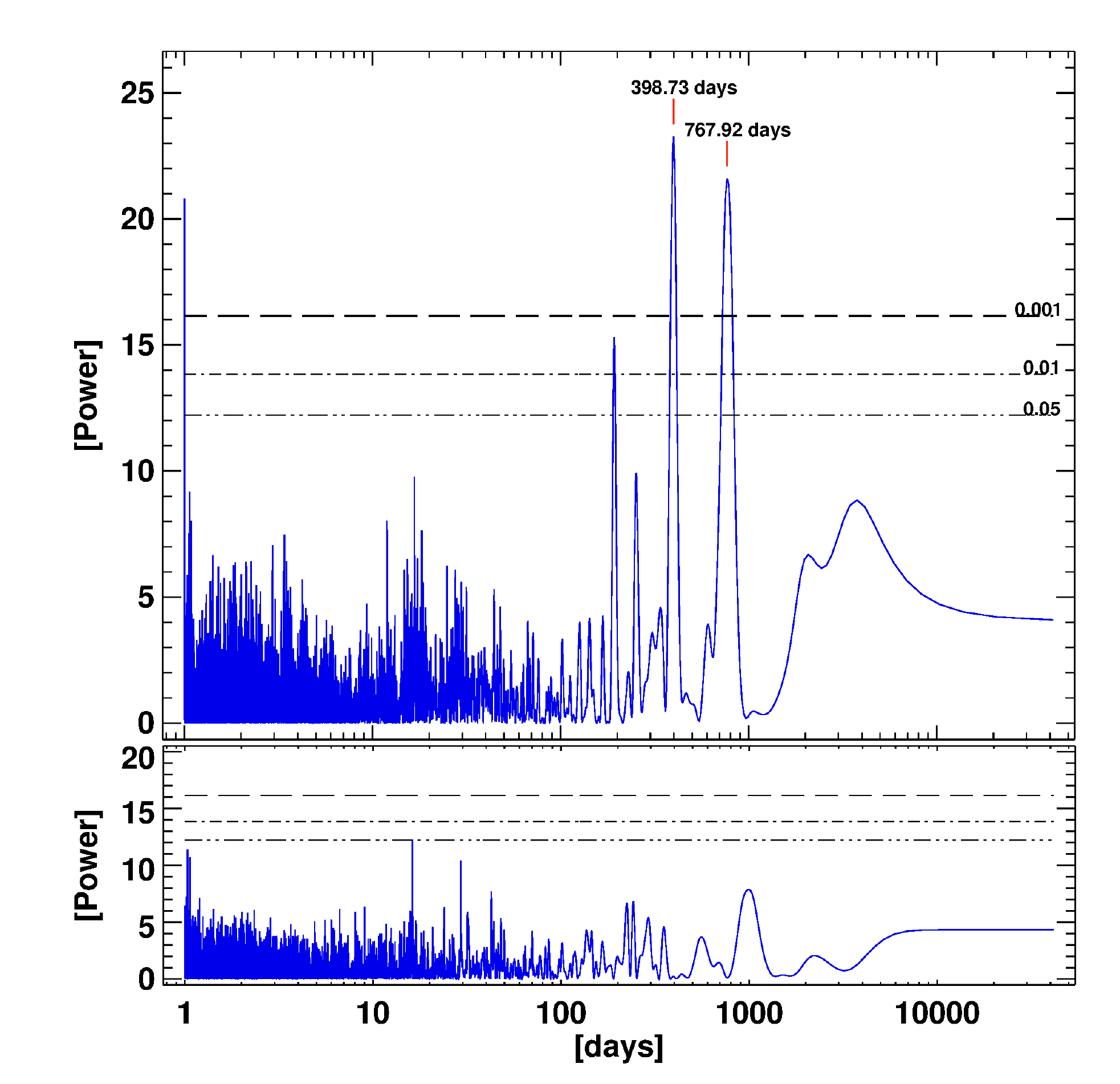}}

\caption{$Top~panel$: the periodogram of the measured radial velocities shows two highly significant peaks
around 399 and 768 days, while the Kepler fit to the data reveals a best fit at periods around 403.5 and
751.9 days. 
$Bottom~panel$: no significant peak is left in the periodogram of the residuals after removing the two periods
from the Keplerian fit.}

\label{FigGam:LOMB}
\end{figure} 

Motivated by these results, our strategy with CRIRES is to test the optical Doppler
data with those from the near-IR regime 
for the objects in our G and K giants sample that clearly exhibit a periodicity consistent 
with one or more substellar companion(s).
If the periodic Doppler signal were indeed caused by a planet, we would expect 
the near-IR radial velocities to follow the best-fit model derived from the optical spectra.

If the radial velocity variations in the optical and in the near-IR are not consistent, 
the reason may be either large stellar spots \citep{Huelamo,Bean} or 
nonradial pulsations that will result in a 
different velocity amplitude at visible and infrared wavelengths \citep{Percy}.
When stellar spots mimic a planetary signal, 
the contrast between the flux coming from the stellar photosphere
and the flux coming from the cooler spot is higher at optical wavelengths and thus has a higher RV amplitude than the near-IR.

\begin{figure}[htp]

\centering

\resizebox{\hsize}{!}{\includegraphics{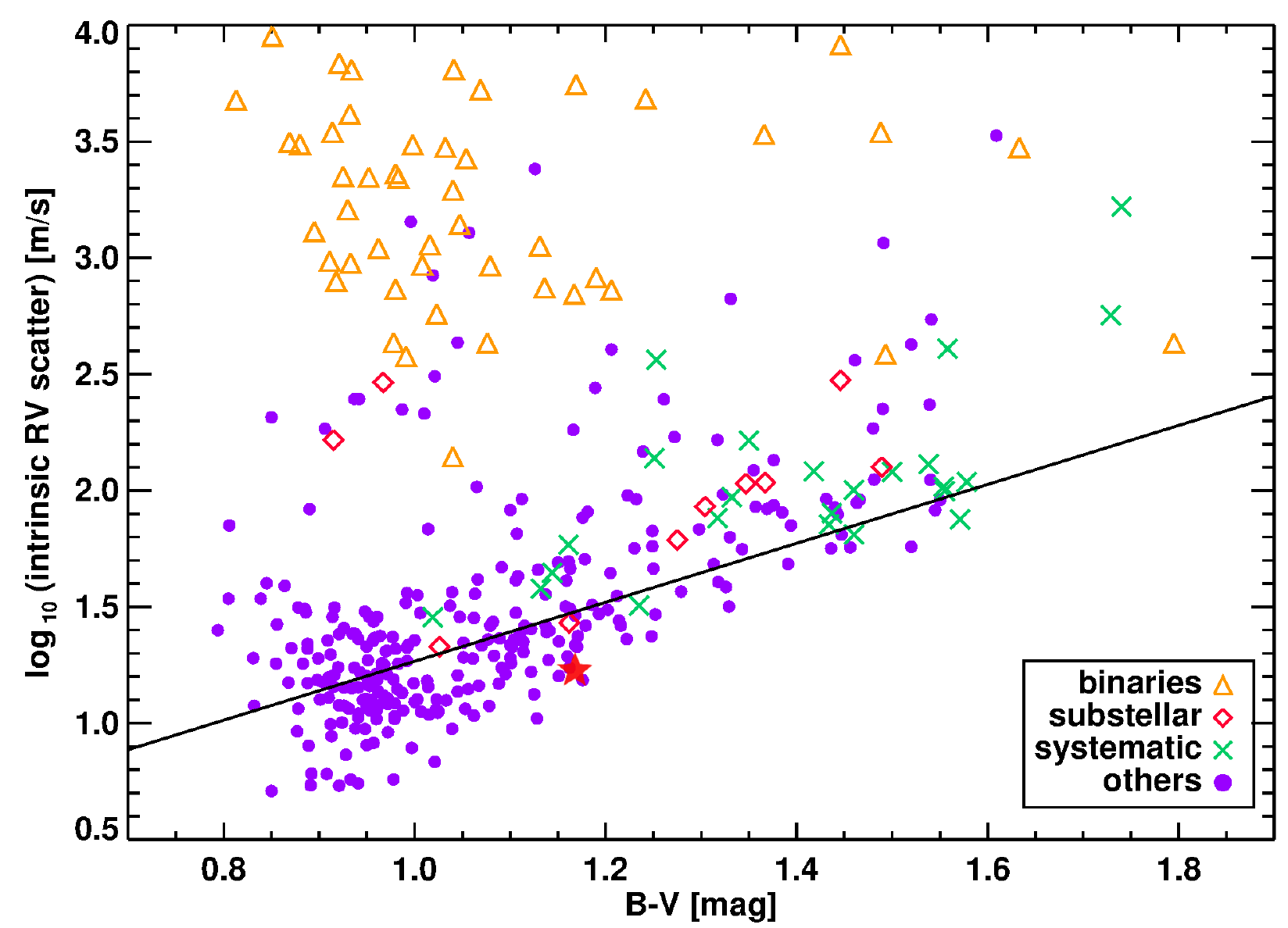}}
\caption{Intrinsic RV scatter observed in our sample of 373 K giants versus $B-V$
color. A clear trend is visible in the sense that redder stars without companions
(${\it circles}$) have larger intrinsic RV
variations. A number of stars lie above the almost linear relation between color and
the logarithm of the scatter. These stars have clearly periodic RV, which indicates that they
harbor substellar or stellar companions.
Stars with non-periodic, but still systematic radial velocities are indicated with green crosses.
The RV scatter of~$\sim$15~m\,s$^{-1}$, for $\eta$~Cet ($red~star$)
derived as the r.m.s.\ around the orbital fit,
is lower than the 25~m\,s$^{-1}$ derived from the linear trend at the star's color index.}

\label{FigGam:Scater}%
\end{figure}


For our observation with CRIRES we decided to adopt an 
observational setup similar to that successfully used by \citet{Figueira}.
We chose a wavelength setting in the $H$ band (36/1/n), with a reference wavelength of 
$\lambda_{\mathrm{ref}}$ = 1594.5~nm.
This particular region was selected by inspecting the Arcturus near-IR spectral atlas from \citet{hinkle95} and 
searching for a good number of stellar as well as telluric lines. The selected spectral region
is characterized by many deep and sharp atmospheric CO$_2$ lines that take
 the role of an always available on-sky gas cell.
To achieve the highest possible precision the spectrograph was used with a resolution of $R$ = 100\,000.
To avoid RV errors related to a nonuniform illumination of the slit,
the observations were made in NoAO mode (without adaptive optics), and nights with poor 
seeing conditions were requested.

Values for the central wavelengths of the telluric CO$_2$ lines were obtained from the HITRAN database
\citep{Rothman}, allowing us to construct an accurate wavelength solution for each detector frame.
The wavelengths of the identified stellar spectral lines were taken from the 
Vienna Atomic Line Database (VALD) \citep{Kupka},
based on the target's $T_{\mathrm{eff}}$ and $\log g$.

Dark, flat, and nonlinearity corrections and the combination of the raw jittered frames in each nodding position
were performed using the standard ESO CRIRES pipeline recipes.
Later, the precise RV was obtained from a cross-correlation \citep{Baranne} of the 
science spectra and the synthetic telluric and stellar mask; this was obtained for each frame and nodding 
position individually.
We estimate the formal error of our CRIRES measurements 
to be on the order of $\sim$~40~m\,s$^{-1}$, based on the $r.m.s.$ dispersion values around
the best fits for all good targets of our CRIRES sample.
This error is probably overestimated, in particular, for the $\eta$~Cet data set.

The full procedure of radial velocity extraction based on the cross-correlation method
will be described in more detail in a follow-up paper (Trifonov et al., in preparation).

The CRIRES observations of $\eta$~Cet were taken with an exposure time of 3 seconds, resulting in a S/N of $\approx$ 300.
Our measured CRIRES radial velocities for $\eta$~Cet are shown together with the data from Lick Observatory
in Fig.~\ref{FigGam:rv}, while the measured values are given in Table~\ref{table:rvlick}.

\begin{table*}[!lht]

  
\resizebox{0.65\textheight}{!}{\begin{minipage}{\textwidth}

\caption{$\eta$~Cet system best fits (Jacobi coordinates).}   
\label{table:orb_par_stable}      
        
\begin{tabular}{ l @{ }r @{ }l @{ }r @{ }l }     

       Orb. Param.& $\eta$~Cet& b & $\eta$~Cet& c \\     
\cline{1-5}\noalign{\vskip 0.5mm}\cline{1-5}
\noalign{\vskip 0.5mm}
 
 Best Keplerian~~~~~~~~~~~~~~~~~~ &~~~ & &\\ \noalign{\vskip 0.4mm} 
 \hline\noalign{\vskip 0.5mm}
   $P$ [days]                         & 403.5 &$\pm$ 1.5    & 751.9 &$\pm$ 3.8 \\  
   $m$        [$M_{\mathrm{Jup}}$]    & 2.55  &$\pm$ 0.13   & 3.32  &$\pm$ 0.18 \\
   $e$                                & 0.13  &$\pm$ 0.05   & 0.1   &$\pm$ 0.06\\
   $M$ [deg]                          & 193.5 &$\pm$ 24.6   & 240.5 &$\pm$ 34.8\\
   $\varpi$ [deg]                     & 250.6 &$\pm$ 20.5   & 67.54 &$\pm$ 5.2 \\   
   $K_1$  [m\,s$^{-1}$]               & 49.7  &             & 52.4  & \\
   $a$ [AU]                           & 1.27  &             & 1.93  & \\  
   r.m.s.~[m\,s$^{-1}]$               & 15.9  \\
   RV$_{\mathrm{offset}}$~[m\,s$^{-1}$]          & $-$0.77  \\
   $\chi_{\mathrm{red}}^2$            &  13.67 & (1.17 with jitter)&& \\
   \noalign{\vskip 0.5mm}
\cline{1-5}\noalign{\vskip 0.5mm}\cline{1-5}

\end{tabular}      
        
  \vspace{0.4cm}      
        
\begin{tabular}{ l @{ }r @{ }l  r @{ }l  p{0.1cm} l @{ }r @{ }l  r @{ }l}     

       Orb. Param.& $\eta$~Cet& b & $\eta$~Cet&c  && Orb. Param.& $\eta$~Cet & b & $\eta$~Cet & c \\     
\cline{1-5}\cline{7-11}
\noalign{\vskip 0.5mm}
\cline{1-5}\cline{7-11}
\noalign{\vskip 0.5mm}
 Best coplanar edge-on & & &&&& 2:1 MMR coplanar edge-on &&&&\\ \noalign{\vskip 0.2mm} 
\cline{1-5}
\cline{7-11}
\noalign{\vskip 0.5mm}
   $P$ [days]                         & 407.5  &$\pm$ 2.67   & 739.9  &$\pm$ 4.8  & & $P$ [days]                      & 407.5  &$\pm$ 2.67    & 744.5  &$\pm$ 3.71 \\  
   $m$ [$M_{\mathrm{Jup}}$]           & 2.55   &$\pm$ 0.16   & 3.26   &$\pm$ 0.17 & & $m$        [$M_{\mathrm{Jup}}$] & 2.54   &$\pm$ 0.16    & 3.28   &$\pm$ 0.19 \\
   $e$                                & 0.12   &$\pm$ 0.05   & 0.08   &$\pm$ 0.04 & & $e$                             & 0.155  &$\pm$ 0.05    & 0.025  &$\pm$ 0.05 \\
   $M$ [deg]                          & 208.2  &$\pm$ 13.7   & 227.8  &$\pm$ 19.6 & & $M$ [deg]                       & 211.1  &$\pm$ 45.33   & 268.0  &$\pm$ 21.33 \\
   $\varpi$ [deg]                     & 245.4  &$\pm$ 9.5    & 68.2   &$\pm$ 22.3 & & $\varpi$ [deg]                  & 244.7  &$\pm$ 31.64   & 32.5   &$\pm$ 32.72\\   
   $K_1$ [m\,s$^{-1}$]                & 49.4   &             & 51.6   &           & & $K_1$ [m\,s$^{-1}$]             & 49.6   &        & 51.7    &  \\
   $a$ [AU]                           & 1.28   &             & 1.91   &           & & $a$ [AU]                        & 1.28   &       & 1.92    &  \\   
   r.m.s.~[m\,s$^{-1}$]               & 15.19  &             &        &           & & r.m.s.~[m\,s$^{-1}$]            & 15.28  &   &   &   \\
   RV$_{\mathrm{offset}}$~[m\,s$^{-1}$] & 0.0    &             &        &           & & RV$_{\mathrm{offset}}$~[m\,s$^{-1}$]       & $-$0.08&   &  &     \\ 
   $\chi_{\mathrm{red}}^2$            &  11.39 &(1.001 with jitter)   &        &  & & $\chi_{\mathrm{red}}^2$         &  11.65 &(1.013 with jitter)       &   &\\
   \noalign{\vskip 0.5mm}
\cline{1-5}\cline{7-11}
\noalign{\vskip 0.5mm}
\cline{1-5}\cline{7-11}
\noalign{\vskip 0.1cm}

 Best coplanar inclined & & &&&&2:1 MMR coplanar inclined \\ \noalign{\vskip 0.2mm} 
\cline{1-5}
\cline{7-11}
\noalign{\vskip 0.5mm}
   $P$ [days]                         & 396.8  &$\pm$ 0.1    & 767.1 &$\pm$ 0.27 & & $P$ [days]                       & 407.3   &$\pm$ 2.1    & 744.3  &$\pm$ 4.1 \\  
   $m$        [$M_{\mathrm{Jup}}$]    & 3.85   &$\pm$ 0.03   & 5.52  &$\pm$ 0.02 & & $m$        [$M_{\mathrm{Jup}}$]  & 2.46    &$\pm$ 0.12   & 3.16   &$\pm$ 0.2 \\
   $e$                                & 0.24   &$\pm$ 0.05   & 0.1   &$\pm$ 0.01 & & $e$                              & 0.17    &$\pm$ 0.05   & 0.02   &$\pm$ 0.03 \\
   $M$ [deg]                          & 163.3  &$\pm$ 0.1    & 78.8  &$\pm$ 0.05 & & $M$ [deg]                        & 208.9   &$\pm$ 16.1   & 262.7  &$\pm$ 41.2\\
   $\varpi$ [deg]                     & 292.2  &$\pm$ 0.1    & 221.3 &$\pm$ 0.1  & & $\varpi$ [deg]                   & 247.2   &$\pm$ 13.5   & 36.67  &$\pm$ 41.1 \\   
   $i$ [deg]                          & 35.5   &             & 35.5  &           & & $i$ [deg]                        & 81.9    &             & 81.9   & \\
   $K_1$ [m\,s$^{-1}$]                & 44.8   &             & 50.2  &           & & $K_1$ [m\,s$^{-1}$]              & 49.8    &             & 51.6  & \\
   $a$ [AU]                           & 1.26   &             & 1.96  &           & & $a$ [AU]                         & 1.28    &             & 1.92  & \\   
   r.m.s.~[m\,s$^{-1}$]               & 14.56  &             &       &           & & r.m.s.~[m\,s$^{-1}$]             & 15.2    &             &   &   \\
   RV$_{\mathrm{offset}}$~[m\,s$^{-1}$]         &$-$3.28 &             &       &           & & RV$_{\mathrm{offset}}$~[m\,s$^{-1}$]       & $-$0.58 &             &   &   \\ 
   $\chi_{\mathrm{red}}^2$            & 10.89  &(0.925 with jitter)  &       &   & & $\chi_{\mathrm{red}}^2$          & 11.84    &(1.04 with jitter)  &   &\\
   \noalign{\vskip 0.5mm}
\cline{1-5}\cline{7-11}
\noalign{\vskip 0.5mm}
\cline{1-5}\cline{7-11}
\noalign{\vskip 0.1cm}

 Best mutually inclined  & & &&&&2:1 MMR mutually inclined  \\ \noalign{\vskip 0.2mm} 
\cline{1-5}
\cline{7-11}
\noalign{\vskip 0.5mm}
   $P$ [days]                         & 404.4   &$\pm$ 2.7    & 748.2 &$\pm$ 6.8   & & $P$ [days]                         & 407.8   &$\pm$ 2.9    & 742.2 &$\pm$ 4.5  \\  
   $m$        [$M_{\mathrm{Jup}}$]    & 5.5     &$\pm$ 1.08   & 7.74  &$\pm$ 1.45  & & $m$        [$M_{\mathrm{Jup}}$]    & 2.45    &$\pm$ 0.12   & 3.14  &$\pm$ 0.17 \\
   $e$                                & 0.06    &$\pm$ 0.06   & 0.05  &$\pm$ 0.07  & & $e$                                & 0.13    &$\pm$ 0.05   & 0.06  &$\pm$ 0.04 \\
   $M$ [deg]                          & 168.4   &$\pm$ 28.1   & 145.3 &$\pm$ 31.1  & & $M$ [deg]                          & 209.0   &$\pm$ 15.4   & 247.2 &$\pm$ 54.5 \\
   $\varpi$ [deg]                     & 283.7   &$\pm$ 23.3   & 138.8 &$\pm$ 70.2  & & $\varpi$ [deg]                     & 246.8   &$\pm$ 12.7   & 51.1  &$\pm$ 52.2 \\   
   $i$ [deg]                          & 151.7   &$\pm$ 24.9   & 155.0 &$\pm$ 34.6  & & $i$ [deg]                          & 88.0    &             & 92.0  & \\ 
   $\Delta\Omega$ [deg]               & 345.4   &$\pm$ 46.4   &       &            & & $\Delta\Omega$ [deg]               & 0.0     & &     &   \\
   $K_1$ [m\,s$^{-1}$]                & 50.2   &             & 52.1 &            & & $K_1$ [m\,s$^{-1}$]                & 47.7    &             & 49.5  &           \\
   $a$ [AU]                           & 1.28    &             & 1.93  &            & & $a$ [AU]                           & 1.29    &             & 1.92  &            \\   
   r.m.s.~[m\,s$^{-1}$]               & 14.61   &             &       &            & & r.m.s.~[m\,s$^{-1}$]               & 15.4    &             &       &            \\
   RV$_{\mathrm{offset}}$~[m\,s$^{-1}$]         & $-$1.89 &             &       &            & & RV$_{\mathrm{offset}}$~[m\,s$^{-1}$]          & $-$0.46 &   &   &   \\ 
   $\chi_{\mathrm{red}}^2$            & 10.90  &(0.95 with jitter)   &       &    & & $\chi_{\mathrm{red}}^2$            & 11.84   &(1.03 with jitter)   &       &   \\
   \noalign{\vskip 0.5mm}
\cline{1-5}\cline{7-11}
\noalign{\vskip 0.5mm}
\cline{1-5}\cline{7-11}

\vspace{-0.1cm} 
\end{tabular}
\end{minipage}}
\end{table*}

\section{Orbital fit}
\label{Orbital fit}

Our measurements for $\eta$ Cet, together with the formal errors
and the best Keplerian and dynamical edge-on coplanar fits to the data, are shown in Fig \ref{FigGam:rv}.  
We used the {\it Systemic~Console} package \citep{Meschiari} for the fitting.

A preliminary test for periodicities with a Lomb-Scargle periodogram shows
two highly significant peaks around 399 and 768 days, suggesting two substellar companions around $\eta$~Cet 
(see Fig.~\ref{FigGam:LOMB}).

The sum of two Keplerian orbits provides a reasonable explanation of the $\eta$~Cet radial velocity data 
(see Fig.~\ref{FigGam:rv}). 
However, the relatively close planetary orbits
and their derived minimum masses raise the question whether this planetary
system suffers from sufficient gravitational perturbations between the bodies that might be detected in the observed data.
For this reason we decided to use Newtonian dynamical fits, applying
the Gragg-Bulirsch-Stoer integration method \citep[B-SM:][]{Press}, built into {\it Systemic}. In this case 
gravitational perturbations that occur between the planets are taken into account in the model.

We used the simulated annealing method \citep[SA:][]{Press} to determine
whether there is more than one $\chi_{\mathrm{red}}^2$ local minimum in the data.
When the global minimum is found, the derived Jacobi orbital elements from the dynamical fit are 
the masses of the planets $m_{b,c}$, the periods $P_{b,c}$, 
eccentricities $e_{b,c}$, longitudes of periastron $\varpi_{b,c}$, and the mean anomaly $M_{b,c}$
(b always denotes the inner planet and c the outer planet).
To explore the statistical and dynamical properties of the fits around the best fit, 
we adopted the systematic grid-search techniques coupled with dynamical fitting. 
This technique is fully described for the HD~82943 two-planet system \citep{Tan2013}.

It is important to note that a good~fit means that the $\chi_{\mathrm{red}}^2$ solution is close to one.
In our case the best edge-on coplanar fit has $\chi_{\mathrm{red}}^2$ = 11.39 
(see \S\ref{Formally best edge-on coplanar fit})
for 118 radial velocity data points, meaning that 
the data are scattered around the fit, and this can indeed be seen in Fig.~\ref{FigGam:rv}.
The reason for that is additional radial velocity stellar~jitter of about
 $\sim$15 m\,s$^{-1}$ that is not taken into account in the weights of each data point. 

This jitter value was determined directly as the r.m.s.\ of the residual deviation around the model.
In fact, this value is close to the bottom envelope of the points in 
Fig.~\ref{FigGam:Scater} for $\eta$~Cet's color index ($B-V$ = 1.16),
which is likely the lowest jitter for $\eta$~Cet.
Based on the long period of our study and the large sample of stars with similar physical characteristics, 
we found that the intrinsic stellar jitter is clearly correlated with the color index $B-V$ of the stars. and 
its value for $\eta$~Cet is typical for other K2 III giants in our Lick sample (see~Fig.~\ref{FigGam:Scater}). 
For $\eta$~Cet we estimated an expected jitter value of 25~m\,s$^{-1}$ 
(see Fig.~\ref{FigGam:Scater}), which is higher than the jitter estimated from the r.m.s.\ of the fit.
It is also known that late G giants \citep{Frandsen, De_Ridder}  and K giants \citep{Barban, Zechmeister2008}
exhibit rapid solar-like $p$-mode oscillations, much more rapid than the typical time 
sampling of our observations, which appear as scatter in our data. 
Using the scaling relation from \citet{Kjeldsen1995},
the stellar oscillations for $\eta$~Cet are estimated to have a period of $\sim$ 0.4 days 
and an amplitude of $\sim$ 12~m\,s$^{-1}$, which  
again agrees well with $\eta$~Cet's RV scatter level around the fit.

We re-assessed the $\chi_{\mathrm{red}}^2$ by quadratically adding the stellar jitter 
to the formal observational errors (3--5 m\,s$^{-1}$) for each radial velocity data point,
which scaled down the $\chi_{\mathrm{red}}^2$ of the best fit close to unity. 
For our edge-on coplanar case $\chi_{\mathrm{red}}^2$ = 11.39 is scaled
down to $\chi_{\mathrm{red}}^2$ = 1.001.
We provide both: the unscaled $\chi_{\mathrm{red}}^2$ value together with the derived stellar jitter 
 and the $\chi_{\mathrm{red}}^2$ value where the average stellar jitter derived above is taken into account.

We estimated the error of the derived orbital parameters 
using two independent methods available as part of the $Console$: bootstrap synthetic data-refitting and
MCMC statistics, which runs multiple MCMC chains in parallel with an adaptive step length.
Both estimators gave similar formal errors for the orbital parameters.
However, the MCMC statistics has been proven to provide better estimates of planetary orbit uncertainties than the 
more robust bootstrap algorithm \citep[e.g.][]{Ford}. 
Therefore, we will use only the MCMC results in this paper.

The nine near-IR Doppler points from CRIRES are overplotted in Fig.~\ref{FigGam:rv},
but were not used for fitting.
We did not consider the CRIRES data because of their
large uncertainties and the 
negligible total weight to the fit, compared with the Lick data. 
Another complication is the radial velocity offset between the two datasets,
which introduces an additional parameter in the $\chi^2$ fitting procedure.
Nevertheless, the CRIRES data points
agree well with the Newtonian fit predictions based on the optical data (see Fig.~\ref{FigGam:rv}),
providing strong evidence for the two-planet hypothesis.

\begin{figure*}[htp]

\centering

\resizebox{\hsize}{!}{\includegraphics[width=18cm]{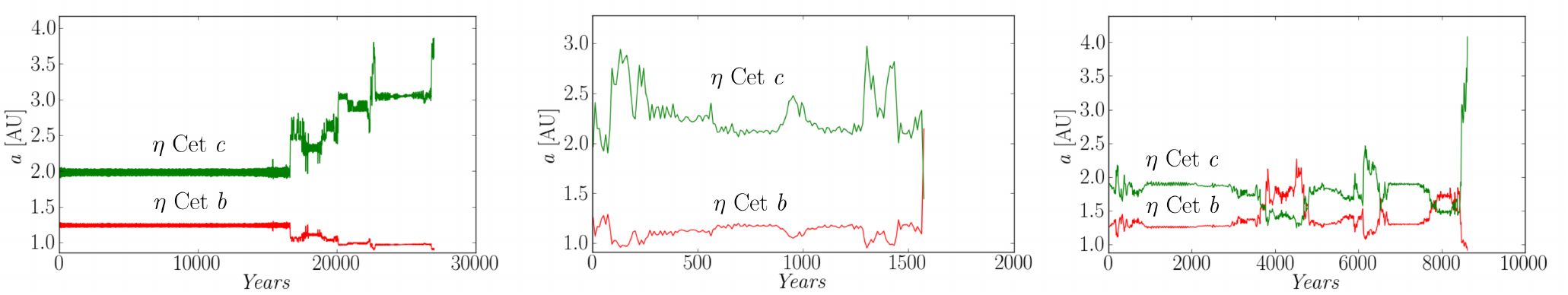}}
\caption{Semimajor axes evolution of the best dynamical fits. Left: the edge-on coplanar fit
remains stable in a 2:1 MMR only for 17\,000 years, when the system starts to show chaotic behavior 
and eventually ejects the outer planet.
The best inclined coplanar (middle) and mutually inclined (right) fits
fail to preserve stability even on very short timescales.
} 

\label{FigGam:unstable}
\end{figure*}

\subsection{Formally best edge-on coplanar fit}
\label{Formally best edge-on coplanar fit}

Assuming an edge-on, co-planar planetary system ($i_{b,c}$ = 90$^\circ$), the global minimum has 
$\chi_{\mathrm{red}}^2$ = 11.39 (1.001 with jitter), which constitutes a significant improvement from the best
two-Keplerian fit with $\chi_{\mathrm{red}}^2$~=~13.67 (1.17 with jitter). This $\chi_{\mathrm{red}}^2$ improvement
indicates that the strong interaction between the two planetary companions is visible in the radial velocity signal
even on short timescales.  
The derived planetary masses are $m_{b}\sin{i_{b}}$ = 2.5~$M_{\mathrm{Jup}}$ and $m_{c}\sin{i_{c}}$ = 3.3~$M_{\mathrm{Jup}}$, 
with periods of $P_{b}$~=~407.5~days and $P_{c}$~=~739.9~days.
The eccentricities are moderate ($e_b$ = 0.12 and $e_c$ = 0.08), and the longitudes of periastron 
suggest an anti-aligned configuration with $\varpi_{b}$ = 245.1$^\circ$ and $\varpi_{c}$ = 68.2$^\circ$, that is, 
 $\varpi_{b}-\varpi_{c}\approx 180^\circ$. 
Orbital parameters for both planets, together with their formal uncertainties, 
are summarized in Table \ref{table:orb_par_stable}.

Dynamical simulations, however, indicate that this fit is stable only for $\lesssim 17\,000$~yr.
After the start of the integrations, the planetary semimajor axes evolution shows a very high perturbation rate with a constant amplitude.
Although the initially derived periods do not suggest any low-order MMR, the average planetary periods
appear to be in a ratio of~2:1 during the first 17\,000 years of orbital evolution,
before the system becomes chaotic and eventually ejects the outer companion. 
This indicates that within the orbital parameter errors, the system might be in a long-term stable~2:1~MMR.
Such stable edge-on cases are discussed in \S\ref{Two planet edge-on coplanar system}  and \S\ref{Coplanar inclined system}.
The evolution of the planetary semimajor axes for this best-fit configuration is illustrated in Fig.~\ref{FigGam:unstable}.

\subsection{Formally best inclined fits} 
\label{Formally best inclined fits} 

We also tested whether our best dynamical fit improved significantly 
by allowing the inclinations with respect to the observer's line of sight (LOS), and the longitudes of the ascending 
nodes of the planets to be $i_{b,c}~\neq$~90$^\circ$  and $\Delta\Omega_{b,c}~=~\Omega_b~-~\Omega_c~\neq$~0, respectively.

The impact of the LOS inclinations on the fits mainly manifests itself through the derived planetary masses.
The mass function is given by
\begin{equation}
  \frac{(m_{p}\sin~i )^3}{(M_{\star}+m_{p})^2} = \frac{P}{2\pi G} K_{\star}^3 \sqrt{(1 - e^2)^3},
\end{equation}

\noindent
where $m_p$ is the planetary mass, $M_{\star}$ the stellar mass, and
$G$ is the universal gravitational constant, while the other parameters come from the orbital model: 
period ($P$), eccentricity ($e$), and radial velocity amplitude ($K_{\star}$). 
It is easy to see that if we take $\sin~i$~$\neq$~1 ($i$ $\neq$ 90$^\circ$), the mass of the planet must increase 
to satisfy the right side of the equation.
However, we note that this general expression is valid only for the simple case of one planet orbiting a star.
Hierarchical two-planet systems and the dependence of the minimum planetary mass on the inclination
are better described in Jacobian coordinates. For more details, we refer to the formalism given in \citet{Lee_m2}.
 
We separated the inclined fits in two different sets, depending on whether
the planets are strictly in a coplanar configuration (yet inclined with respect to the LOS),
or if an additional mutual inclination angle between the planetary orbits is allowed 
(i.e.,\ $i_b~-~i_c~\neq$~0$^\circ$ and $\Omega_{b}$~-~$\Omega_{c}~\neq$~0$^\circ$).

For the inclined co-planar test we set $i_{b}$~=~$i_{c}$, but $\sin{i_{b,c}}$ $\leq$ 1,
and we fixed the longitudes of ascending nodes to $\Omega_{b}$~=~$\Omega_{c}$~=~0$^\circ$.
In the second test the inclinations of both planets were fitted as independent parameters, 
allowing mutually inclined orbits. However, $i_b$ and $i_c$ were restricted to not exceed the
$\sin{i_{b,c}}$ = 0.42 ($i_{b,c}$ = 90$^\circ ~ \pm$65$^\circ$) limit, where the planetary masses will become very large.
Moreover, as discussed in \citet{Laughlin} and in \citet{Bean2}, the mutual inclination ($\varPhi_{b,c}$)
of two orbits depends not only on the inclinations $i_b$ and $i_c$, but also on the longitudes of the ascending nodes 
$\Omega_b$ and $\Omega_c$: 
\vspace{-0.1cm}
\begin{equation}
\cos~\varPhi_{b,c} = \cos~i_{b}~\cos~i_{c} + \sin~i_{b}~\sin~i_{c}~\cos~(\Omega_{b}-\Omega_{c}).
\end{equation}

The longitudes of the ascending node, $\Omega_b$ and $\Omega_c$, are not restricted and 
thus can vary in the range from 0 to 2$\pi$.
The broad range of $i_{b,c}$ and $\Omega_{b,c}$ might lead to very high mutual inclinations, but in general $\varPhi_{b,c}$~was
restricted to 50$^\circ$, although this limit was never reached by the fitting algorithm.

For the coplanar inclined case, the minimum appears to be $\chi_{\mathrm{red}}^2$ = 10.89,
while adding the same stellar jitter as above to the data used for the coplanar fit gives $\chi_{\mathrm{red}}^2$ = 0.925. 
Both planets have orbits with relatively high inclinations with respect to the LOS ($\it{i_{b,c}}$~=~35.5$^\circ$).
The planetary masses are $m_{b}$ = 3.85 $M_{\mathrm{Jup}}$ and $m_{c}$ = 5.52 $M_{\mathrm{Jup}}$, 
and the planetary periods are closer to the 2:1 ratio: $P_{b}$ = 396.8 days and $P_{c}$ = 767.1 days.

The derived mutually inclined best fit has $\chi_{\mathrm{red}}^2$~=~10.90, ($\chi_{\mathrm{red}}^2$~=~0.95).
This fit also has high planetary inclinations and thus the planetary masses are much more massive: 
 $m_{b}$ = 5.50 $M_{\mathrm{Jup}}$ and $m_{c}$ = 7.74 $M_{\mathrm{Jup}}$, while the 
 periods are $P_{b}$ = 404.4 days and $P_{c}$ = 748.2 days.
Orbital parameters and the associated errors for the inclined fits are summarized in Table \ref{table:orb_par_stable}.
An F-test shows that the probability that the three additional fitting 
parameters significantly improve the model is $\sim$ 90\%.

Dynamical simulations based on the inclined fits show that these solutions cannot 
even preserve stability on very short timescales.
The large planetary masses in those cases and the higher interaction rate
make these systems much more fragile than the edge-on coplanar system.
The best inclined co-planar fit appears to be very unstable and leads to planetary collision in less than 1600 years.
The best mutually inclined fit is chaotic from the very beginning of
the integrations. During the simulations
the planets exchange their positions in the system 
until the outer planet is ejected after $\sim$ 9000 years.
The semimajor axes evolution for those systems is illustrated in Fig.~\ref{FigGam:unstable}.

\begin{figure*}[htp]

\centering
 
\resizebox{\hsize}{!}{\includegraphics{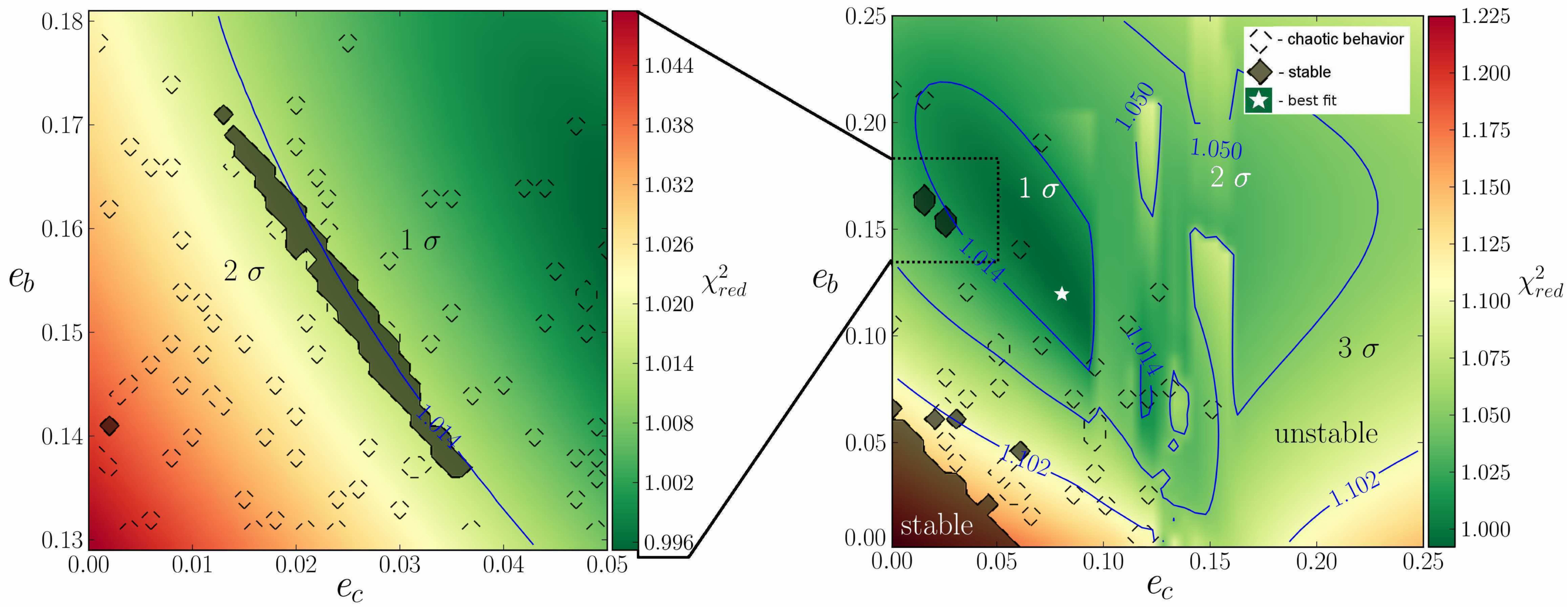}}
 
\caption{ Right: edge-on coplanar $\chi_{\mathrm{red}}^2$ grid with jitter included. The eccentricities
of both planets are varied in the range from 0.001 to 0.251 with steps of 0.005, while the 
other orbital parameters and the zero-point offset were fitted until the $\chi_{\mathrm{red}}^2$ minimum is achieved. 
The solid black contours indicate the stable fits, while
the dashed contours indicate fits where the system survives the dynamical tests, but with $chaotic$ scattering behavior.
While the best dynamical fit is unstable ($white~star$), we found two stability islands where 
long-term ($10^8$ yr) stability is achieved.
With a moderate $e_{b}$, at the 1$\sigma$ border (blue contours), a stable 2:1 MMR region exists,  
and at lower eccentricities a broad stability region can be seen at more than 3$\sigma$ from the best fit,
without showing any signs of a low-order MMR.
Left: higher resolution zoom of the stable 2:1 resonant region. 
}

\label{FigGam:2}%
\end{figure*}   

\section{Stability tests}
\label{Stability tests} 

\subsection{Numerical setup} 
\label{Numerical setup}

For testing the stability of the $\eta$~Cet planetary system we used the {\it Mercury} N-body 
simulator \citep{Chambers}, and the {\it SyMBA} integrator \citep{Duncan}.
Both packages have been designed to calculate the orbital evolution of 
objects moving in the gravitational field of a much more massive central body, as in the case of extrasolar planetary systems.
We used {\it Mercury} as our primary program and {\it SyMBA} to double-check the obtained results.
All dynamical simulations were run using the hybrid sympletic/Bulirsch-Stoer algorithm, which is able to 
compute close encounters between the planets if they occur during the orbital evolution. 
The orbital parameters for the integrations are taken directly from
high-density $\chi_{\mathrm{red}}^2$ grids (see \S\ref{Two planet edge-on coplanar system}, 
\S\ref{Coplanar inclined system}, \S\ref{Mutually inclined system}  and
\S\ref{Stellar mass impact on the stability analysis})
 ($\sim$~120\,000~fits). Our goal is to check the permitted stability regions 
 for the $\eta$~Cet planetary system and to constrain the orbital parameters by requiring stability.

The orbital parameter input for the integrations are in {\it astrocentric} format: mean anomaly $M$,
semimajor axis $a$, eccentricity $e$, argument of periastron $\omega$,
orbital inclination  $i$, longitude of the ascending node $\Omega$, and absolute planetary mass derived from fit 
(dependent on the LOS inclination).
The argument of periastron is $\omega$~=~$\varpi$~-~$\Omega$, 
and for an edge-on or co-planar configuration $\Omega$ is undefined and thus $\omega$~=~$\varpi$.
From the orbital period $P$ and assuming $m_{b,c}\sin~i_{b.c} \ll M_{\star}$,
the semimajor axes $a_{b,c}$ are calculated from the general form for the two-body problem:

\begin{equation}
\it{a_{b,c}} =  \Biggl({\frac{G M_{\star}P_{b,c}^2}{4\pi^2}}\Biggr)^{1/3}.
\end{equation}

\noindent
Another input parameter is the {\it Hill radius}, which indicates the~maximum distance from the body that 
constitutes a close encounter.
A Hill radius approximation \citep{Hamilton1992} is calculated from

\begin{equation}
\it{r_{b,c}}\approx\it{a_{b,c}}(1 - \it{e_{b,c}}) \Biggl(\frac{m_{b,c}}{3M_{\star}}\Biggr)^{1/3}.
\end{equation}

All simulations were started from JD = 2451745.994, the epoch when the first RV observation of $\eta$~Cet was taken, and then
integrated for $10^5$~years.
This timescale was chosen carefully to minimize CPU resources, while still allowing
a detailed study of the system's evolution and stability. 
When a test system survived this period,
we tested whether the system remains stable over a longer period of time by extending 
the integration time to $10^8$~years for the edge-on coplanar fits (\S\ref{Two planet edge-on coplanar system}), 
and to 2x10$^6$~years for inclined configurations
(\S\ref{Coplanar inclined system} and \S\ref{Mutually inclined system}).
On the other hand, the simulations were interrupted
in case of collisions between the bodies involved in the test, or ejection of one of the planets. 
The typical time step we used for each dynamical integration was equal to 
eight days, while the output interval from the integrations was set to one year. 
We defined an ejection as one of the planet's semimajor axes exceeding 5 AU 
during the integration time.

In some cases none of the planets was ejected from the system and no planet-planet or planet-star collisions occurred, but
 because of close encounters, their eccentricities became very high and the semimajor axes
showed single or multiple time planetary scattering to a different semimajor axis within the 5~AU limit.
These systems showed an unpredictable behavior, and we classified them as $chaotic$,
 even though they may not satisfy the technical definition of chaos.

We defined a system to be stable if 
the planetary semimajor axes remained within 0.2~AU from the 
semimajor axes values at the beginning of the simulation during the maximum integration time.
This stability criterion provided us with a very fast and accurate estimate of the dynamical behavior of the system,
and clearly distinguished the stable from the chaotic and the unstable configurations.
In this paper we do not discuss the scattering (chaotic) configurations, but instead focus on the
configurations that we qualify as stable.

\begin{figure*}[!ht]
\begin{center}$
\begin{array}{ccc}
\includegraphics[width=2.3in]{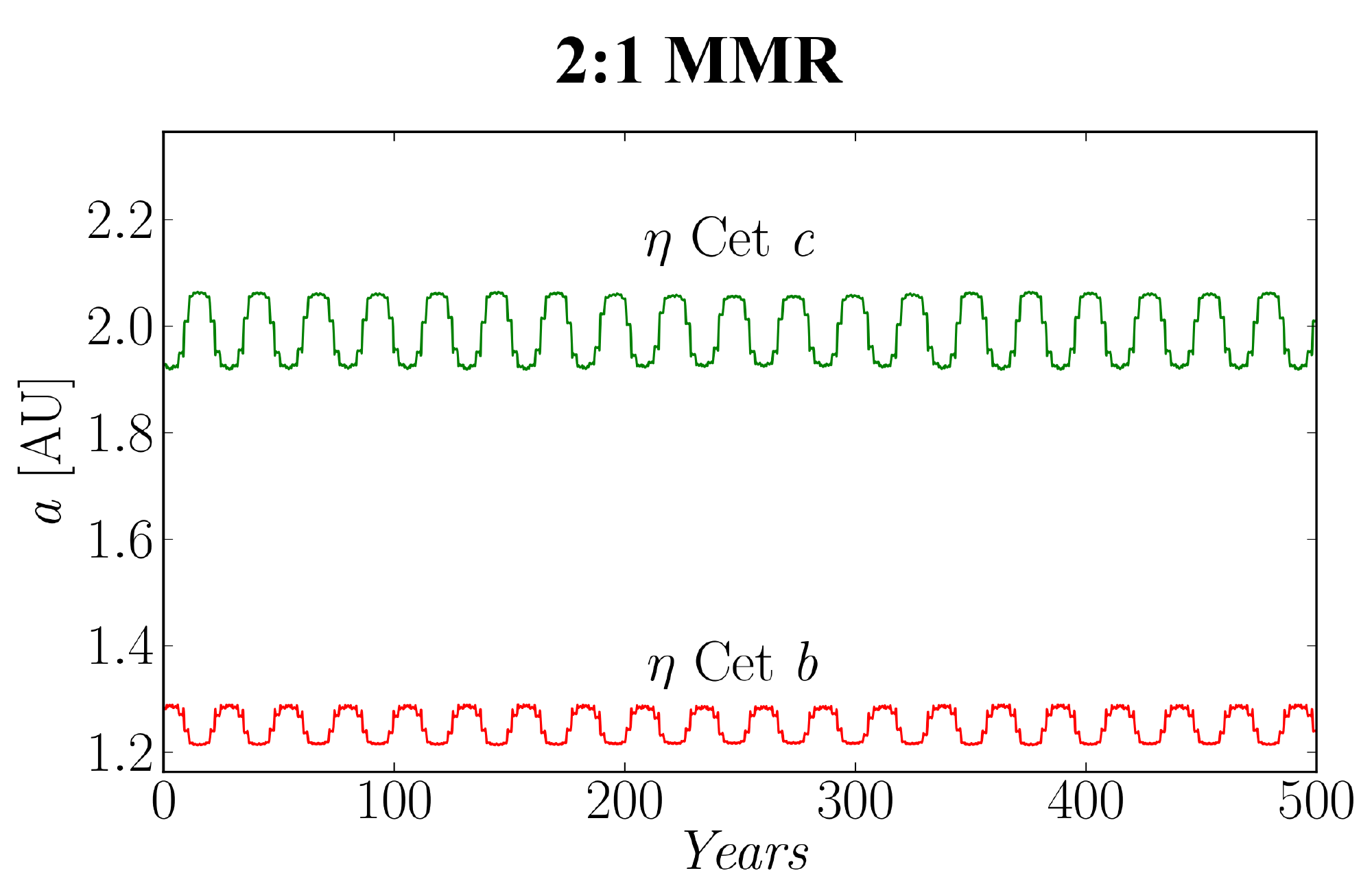} &
\includegraphics[width=2.3in]{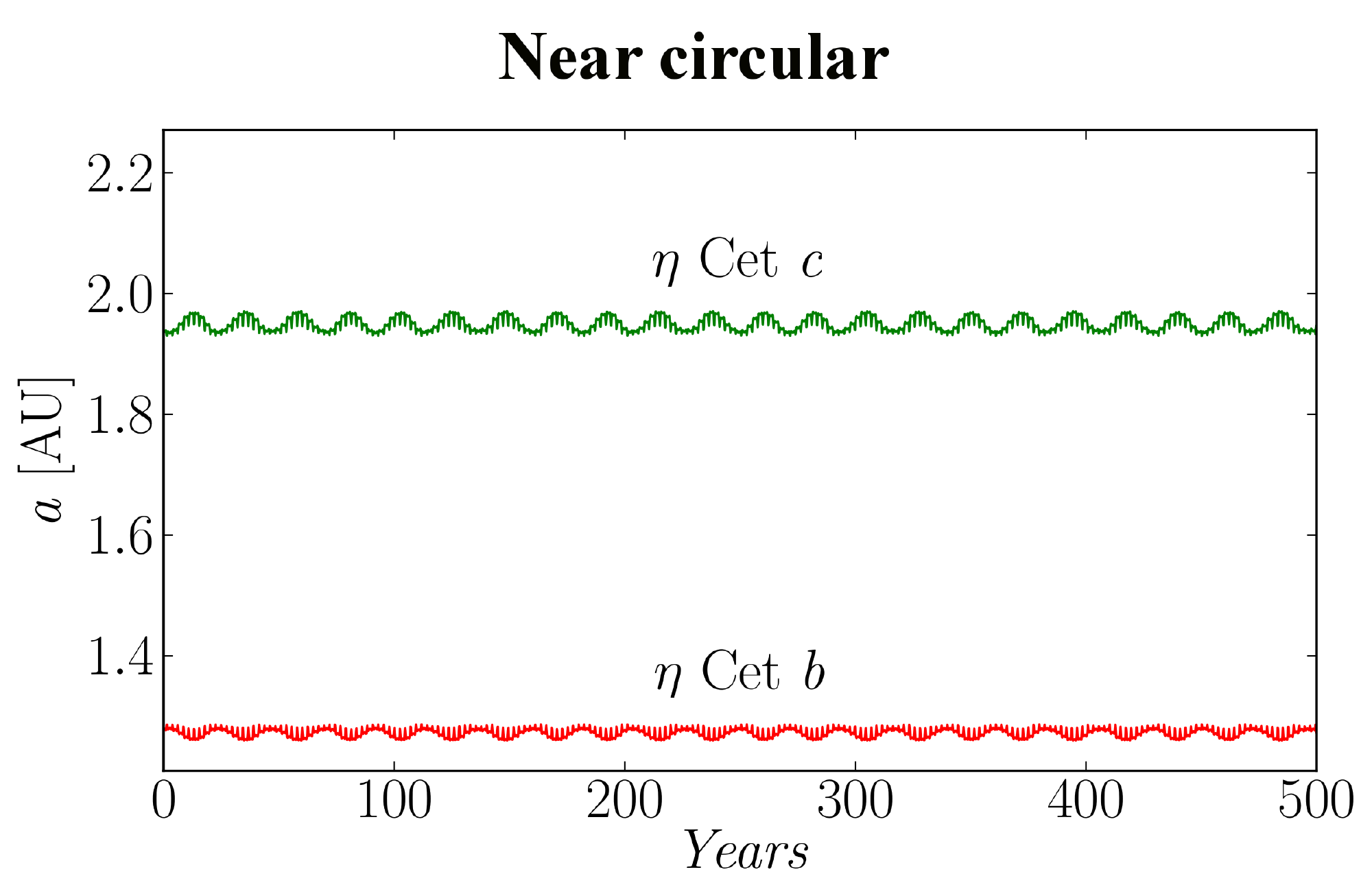} &
\includegraphics[width=2.3in]{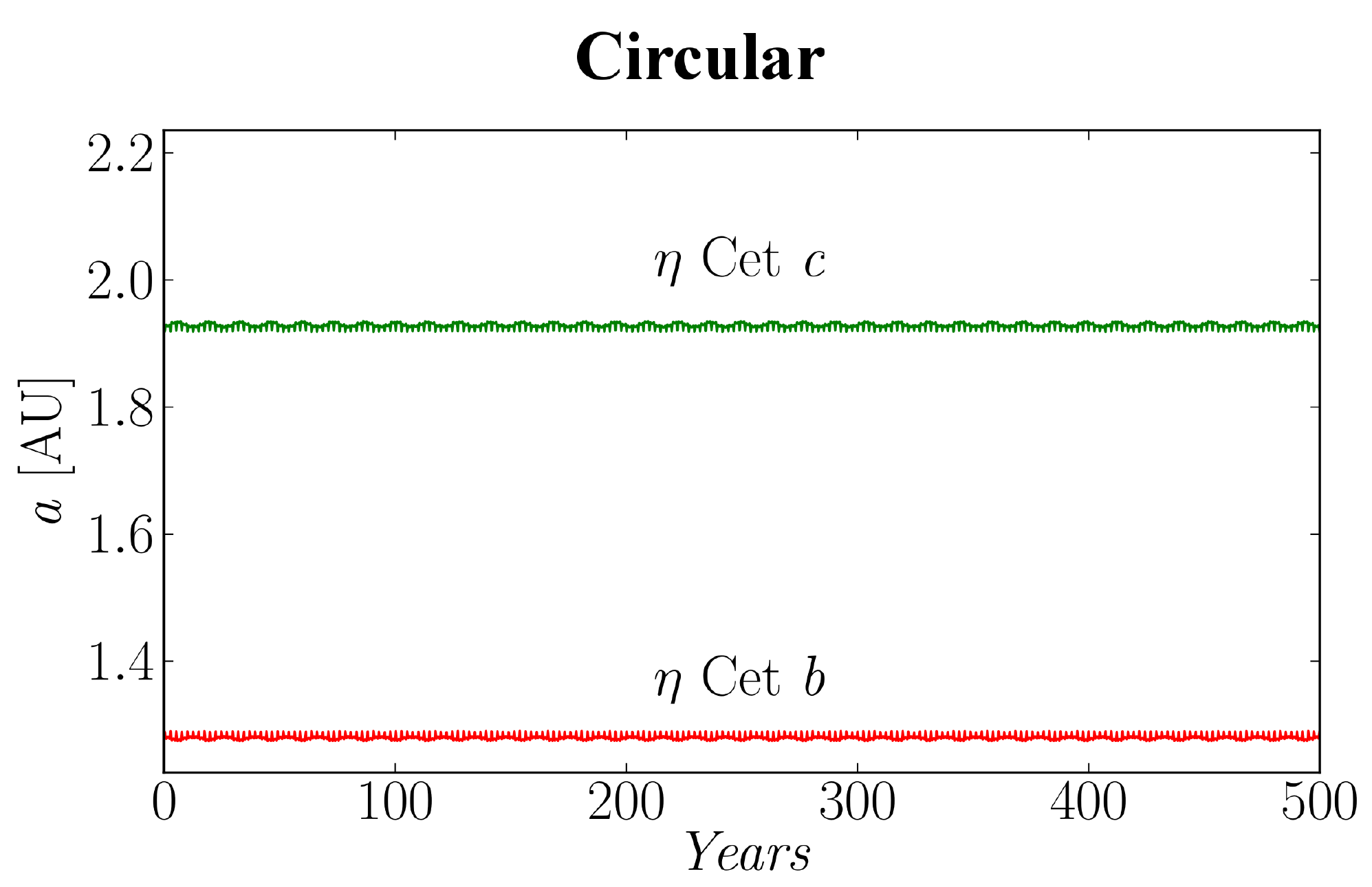}  \\
\includegraphics[width=2.3in]{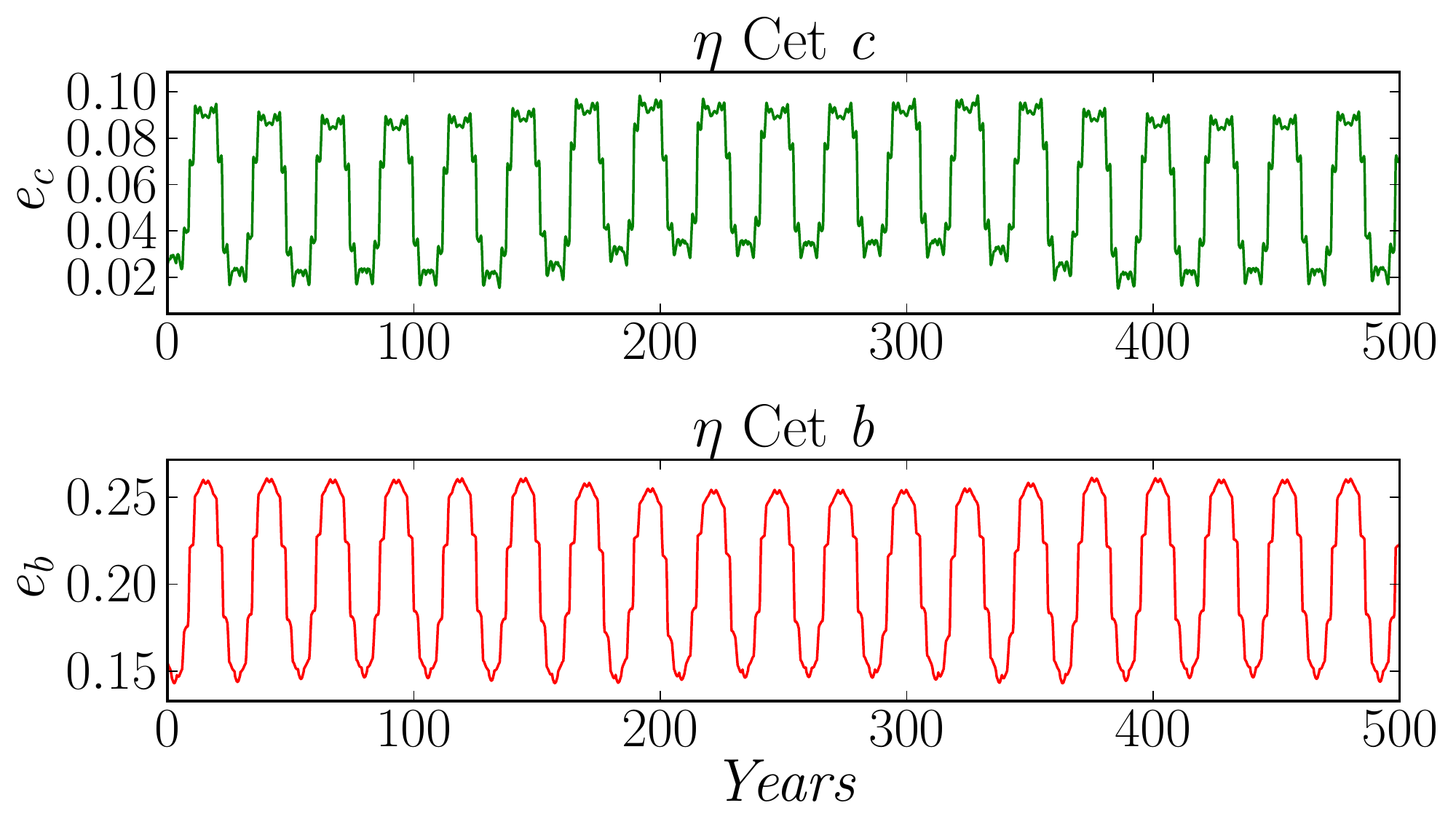} &
\includegraphics[width=2.3in]{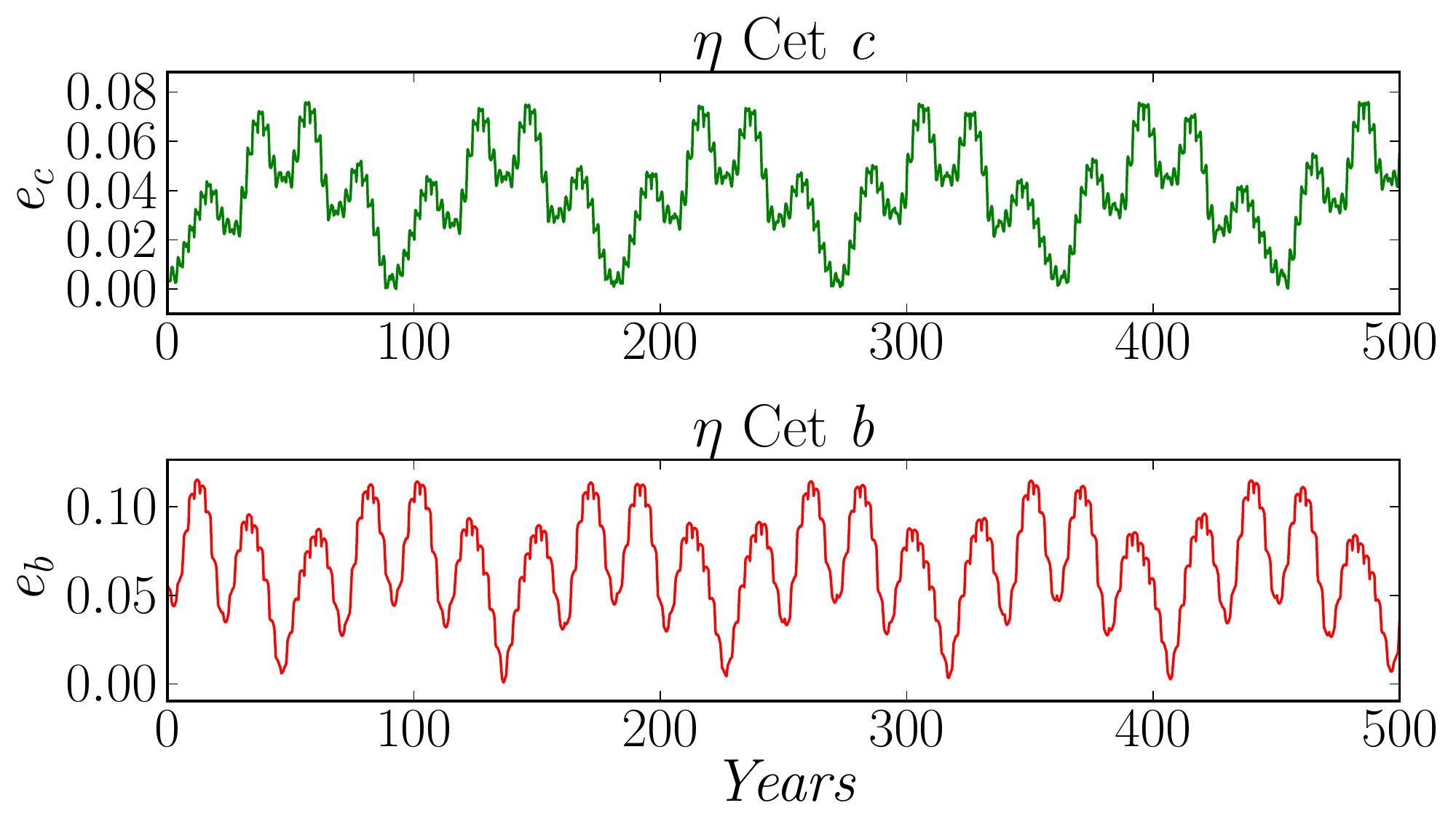} &
\includegraphics[width=2.3in]{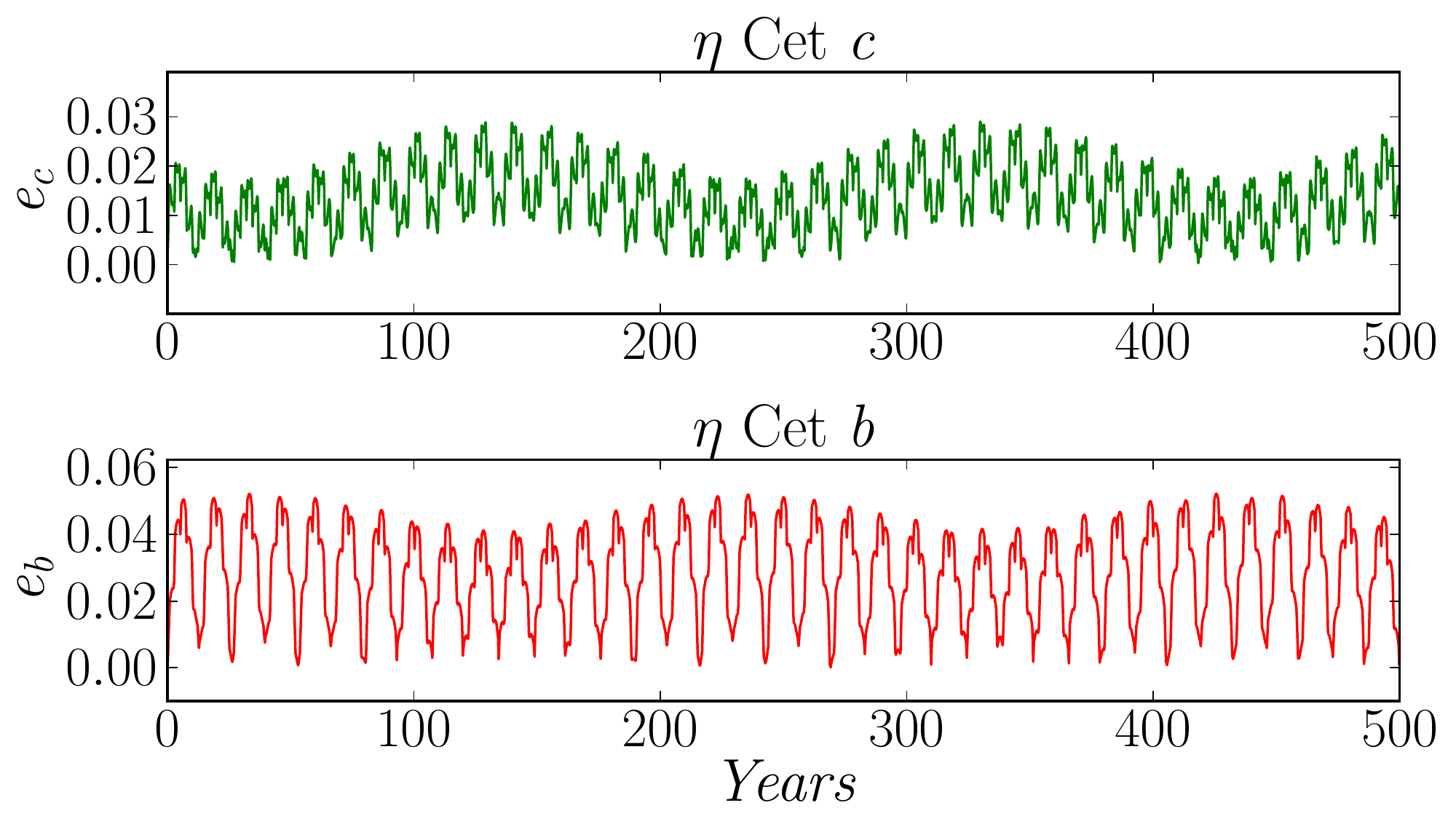}  \\
\includegraphics[width=2.3in]{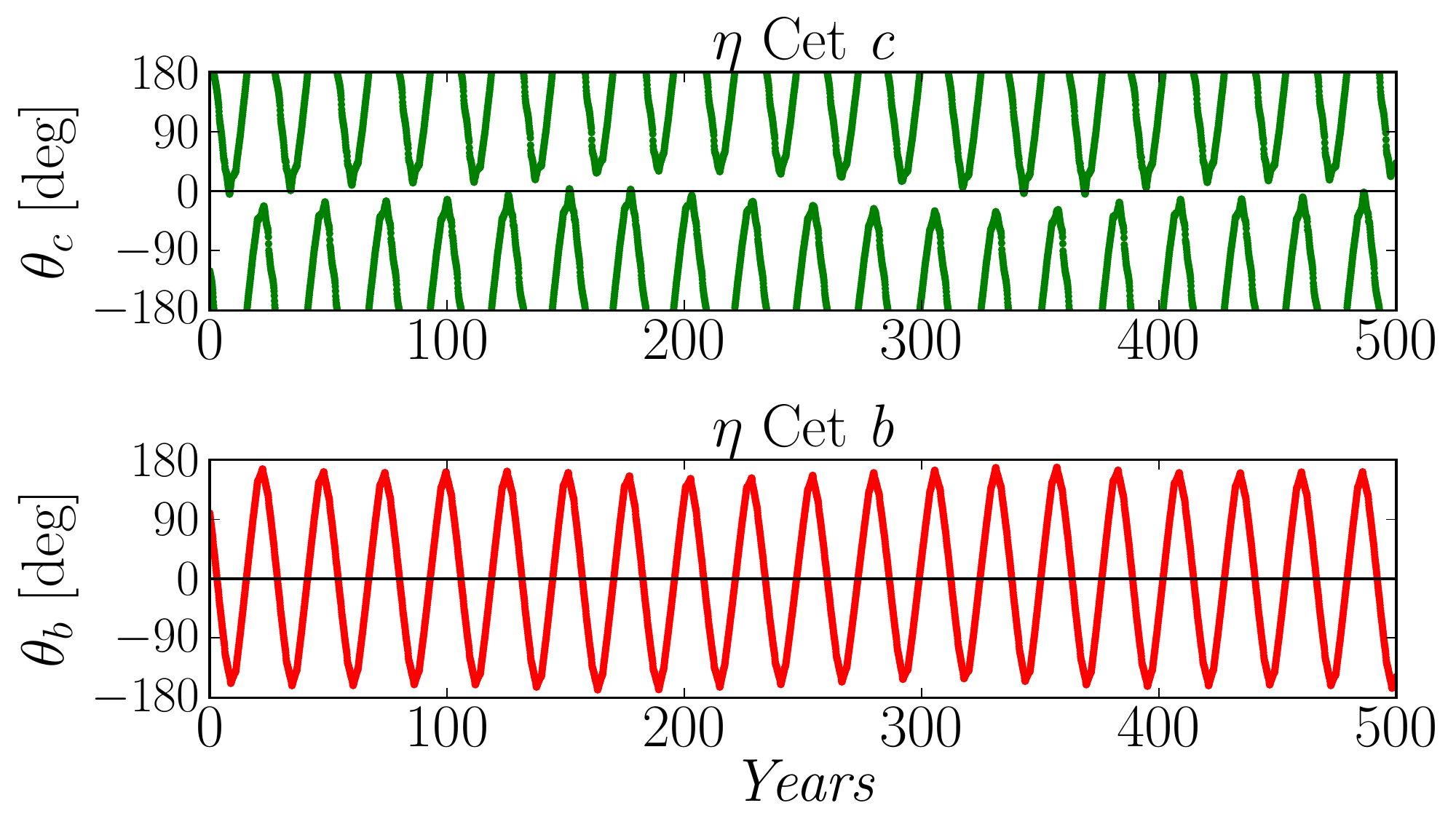} &  
\includegraphics[width=2.3in]{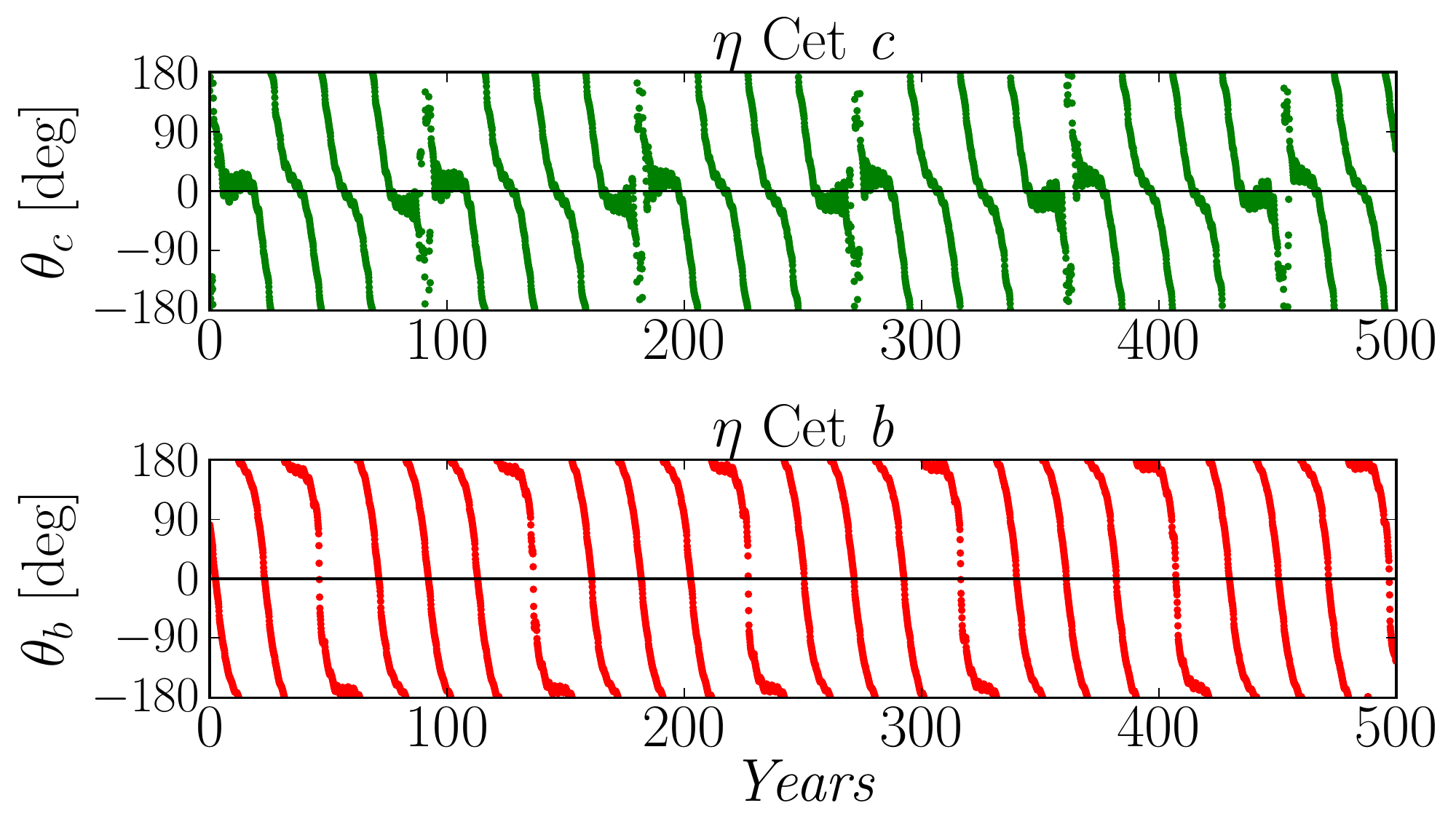} &
\includegraphics[width=2.3in]{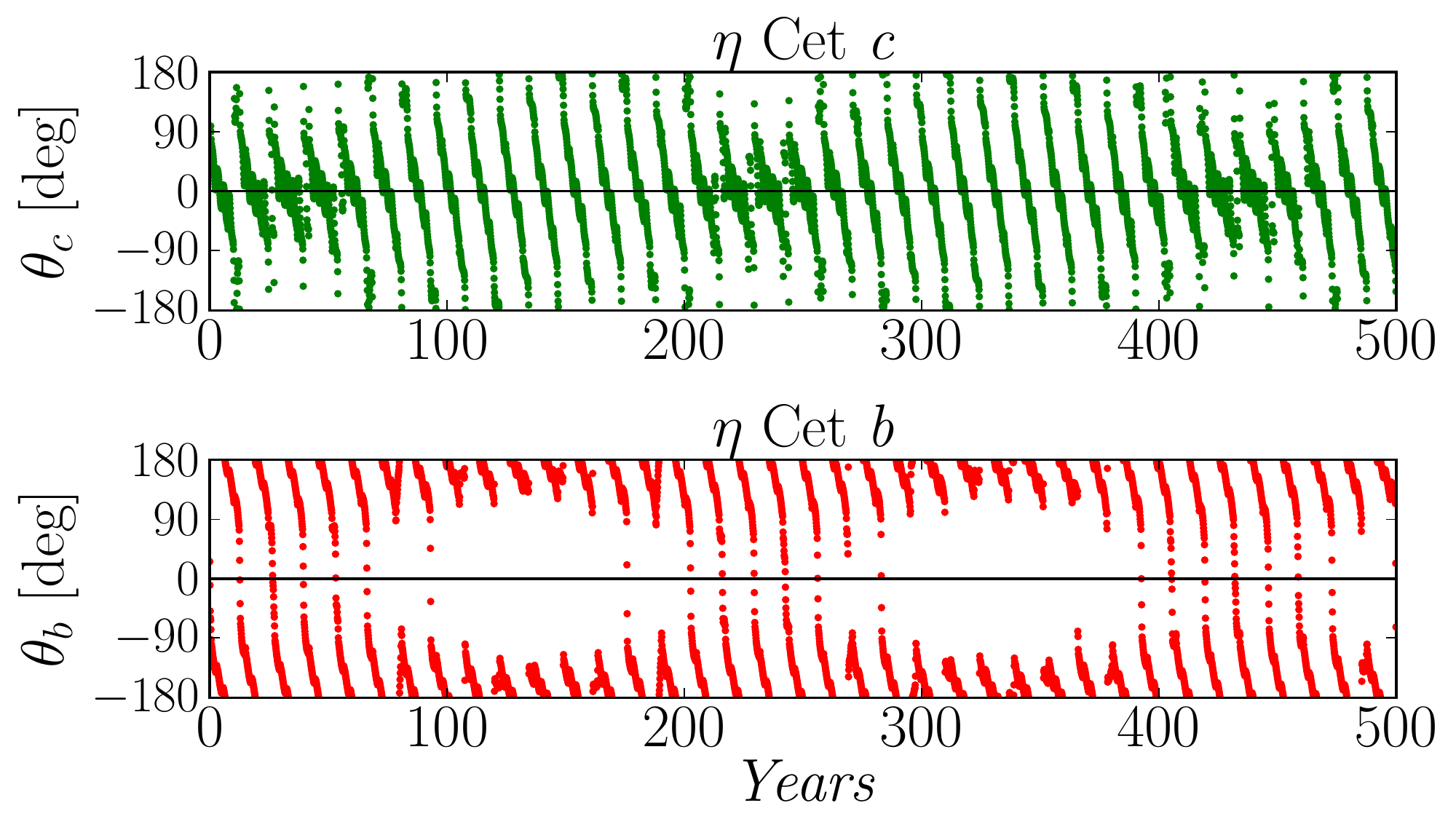} \\
\includegraphics[width=2.3in]{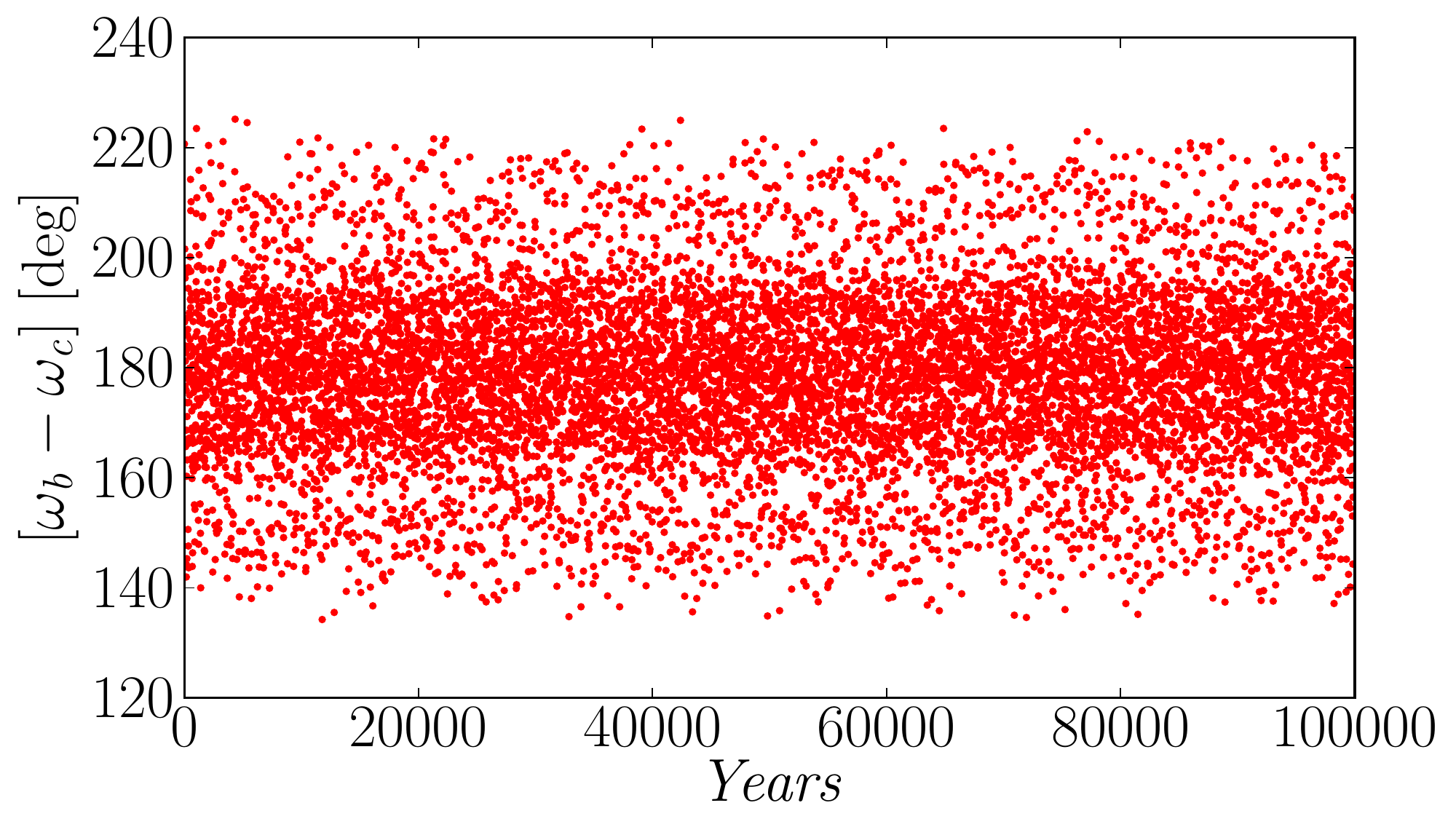} & 
\includegraphics[width=2.3in]{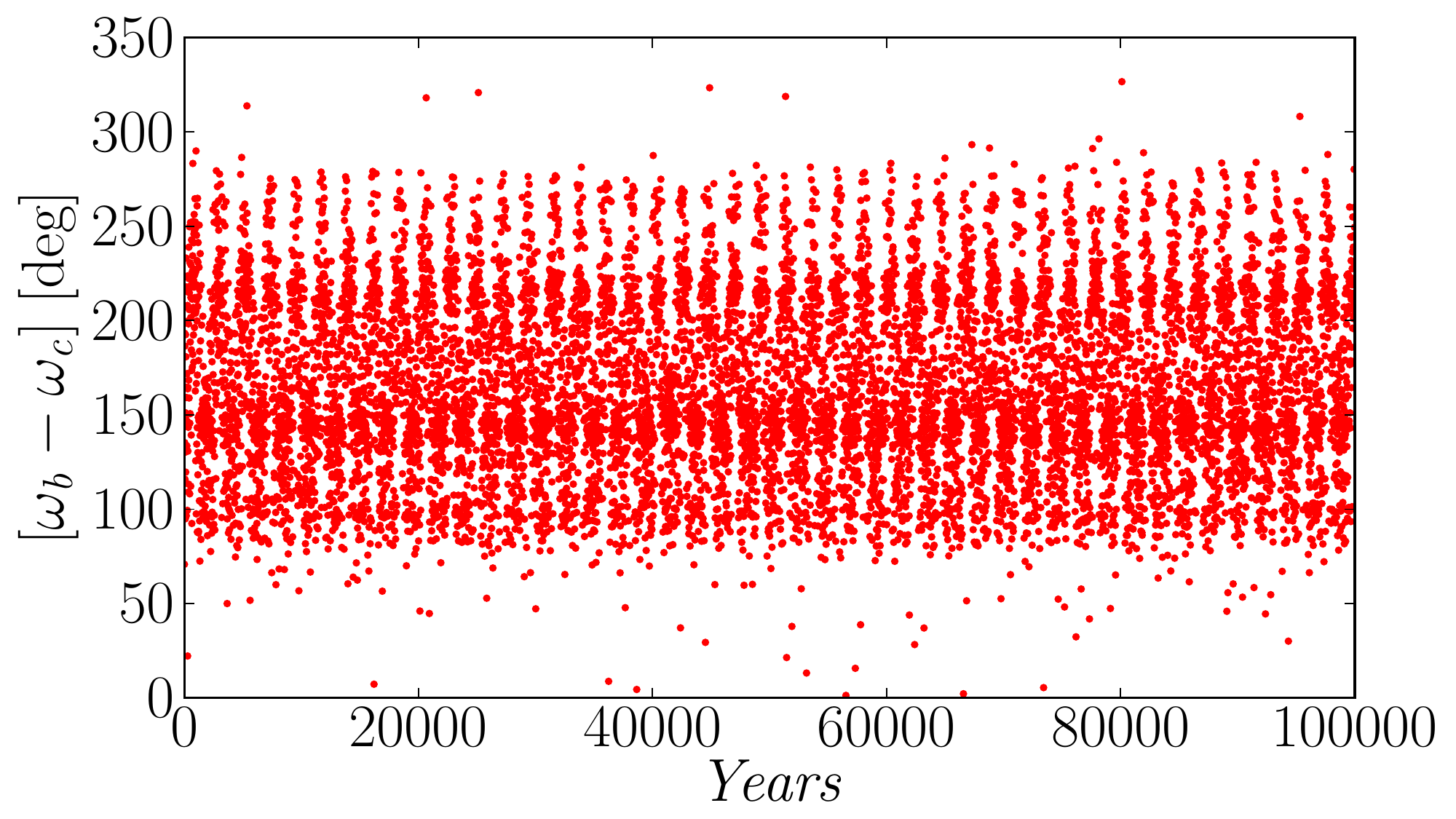} &
\includegraphics[width=2.3in]{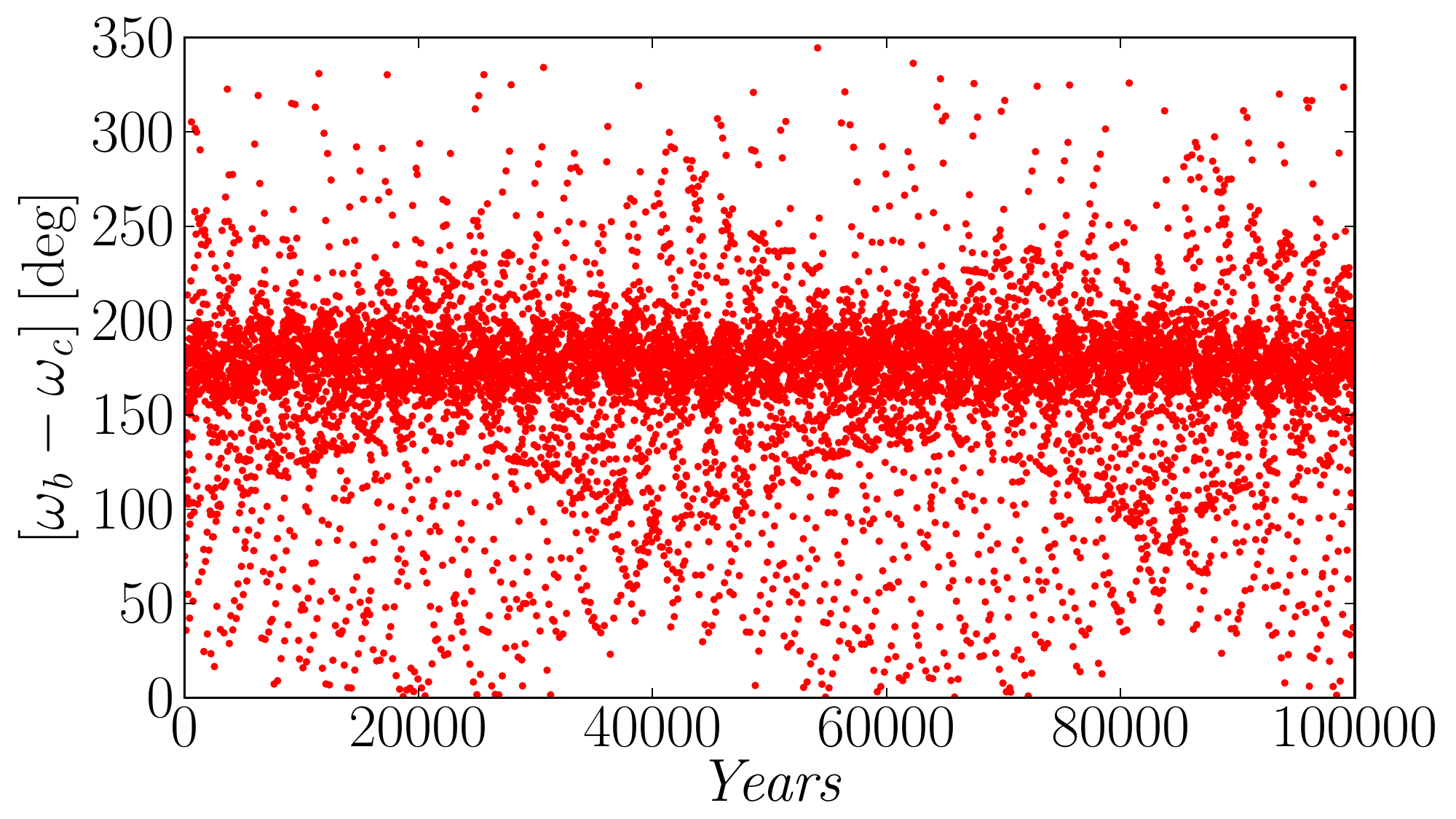} \\
\end{array}$
\includegraphics[width=7.2in]{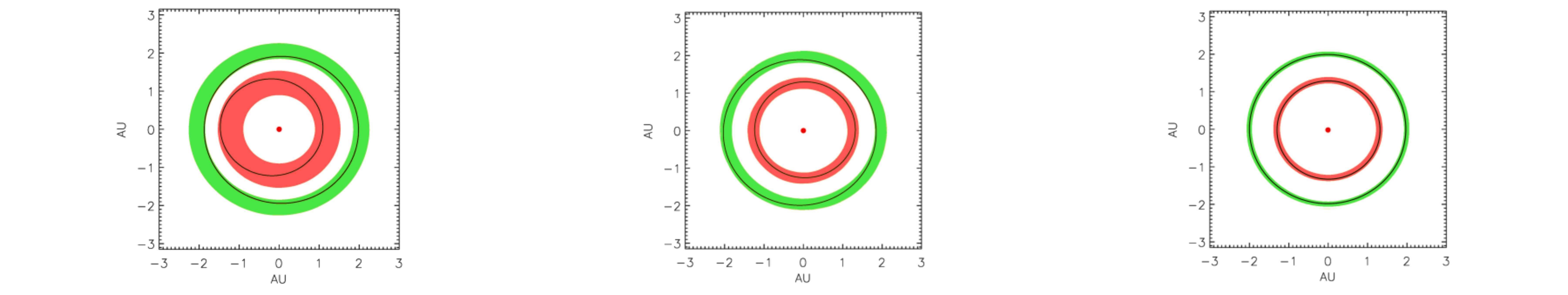} \\
 
\end{center}
\caption{Evolution of the orbital parameters for three different fits, stable for at least 10$^8$ years.
The best 2:1 MMR fit (left panels), the best stable fit from the low-eccentricity region (middle panels), and
the fit with the most circular orbits (right panels).
In the 2:1 MMR fit the gravitational perturbation between the planets is much larger than in the other two cases.
It is easy to see that eccentricities for given epochs can be much larger than their values at the initial epoch of the integration. 
For convenience the evolution of the semimajor axes, eccentricities, 
the resonant angles (third row) and the $\Delta \omega_{b,c}$ are given for 500 and 10$^5$ years, respectively. 
The bottom row gives the orbital precession region, the sum of orbits for each integration output.}
\label{FigGam:1}
\end{figure*}

\begin{figure*}[htp]

\centering

\resizebox{\hsize}{!}{\includegraphics[width=18cm]{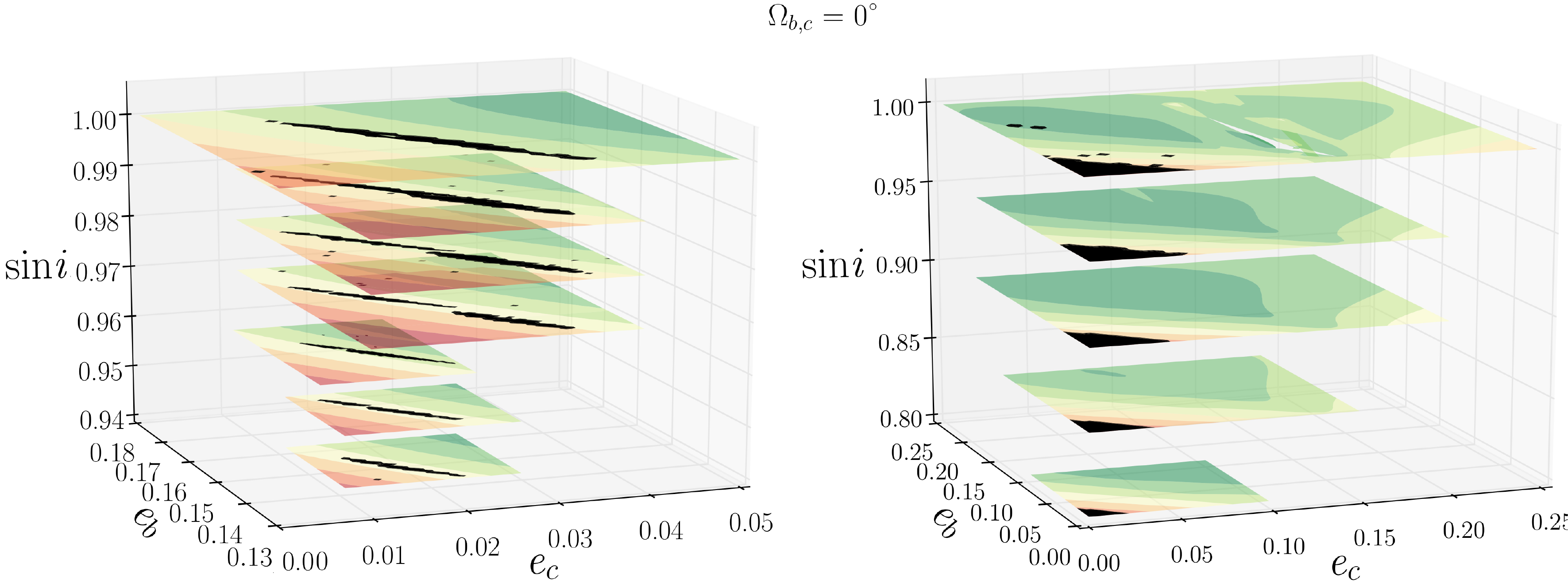}}
\caption{Coplanar inclined grids illustrate the stability dependence on $m_{b,c} \sin{i_{b,c}}$.
Color maps are the same as in Fig.~\ref{FigGam:2}, with the difference that for clarity, only the stable regions are shown 
(black).
The top layer shows the grids from Fig.~\ref{FigGam:2}, where $\sin{i_{b,c}}$~=~1 ($i_{b,c}$ = 90$^\circ$). Decreasing the 
inclination leads to smaller near-circular and 2:1 MMR stability regions. The resonant region shrinks and
moves in the $(e_b, e_c)$ plane, and it completely vanishes when $\sin{i}$ $\leq$ 0.93.
The stable island at low eccentricities vanishes at $\sin{i}$ $\leq$ 0.75, 
when even the most circular ($e_{b,c}$ = 0.001) fit is unstable.
}

\label{FigGam:4}
\end{figure*}   

\subsection{Two-planet edge-on coplanar system} 
\label{Two planet edge-on coplanar system} 
 
The instability of the best fits motivated us to start a high-density $\chi_{\mathrm{red}}^2$ grid search  
 to understand the possible sets of orbital configurations for the $\eta$~Cet planetary system.
To construct these grids we used only scaled $\chi_{\mathrm{red}}^2$ fits
with stellar jitter quadratically added, so that $\chi_{\mathrm{red}}^2$ is close to unity. 
Later, we tested each individual set for stability, transforming these grids to effective stability maps.

In the edge-on coplanar two dimensional grid we varied the eccentricities of the planets 
from $e_{b,c}$ = 0.001 (to have access to $\varpi_{b,c}$) to 0.251 with steps of 0.005 (50x50 dynamical fits),
while the remaining orbital parameters in the model ($m_{b,c}\sin~i_{b,c}$, $P_{b,c}$, $M_{b,c}$, $\varpi_{b,c}$
and the RV offset) were fitted until the best possible solution to the data was achieved. 
The resulting $\chi_{\mathrm{red}}^2$~grid is smoothed with bilinear interpolation between each
grid pixel, and 1,~2~and~3$\sigma$ confidence levels (based on $\Delta\chi_{\mathrm{red}}^2$ from the minimum)
are shown (see Fig.~\ref{FigGam:2}).
The grid itself shows that very reasonable $\chi_{\mathrm{red}}^2$ fits can be found in a
broad range of eccentricities, with a tendency toward lower $\chi_{\mathrm{red}}^2$ values in higher and moderate 
eccentricities, and slightly poorer fits are found for the near-circular orbits. 
However, our dynamical test of the edge-on coplanar grid illustrated in Fig.~\ref{FigGam:2} 
shows that the vast majority of these fits are unstable. The exceptions are 
a few isolated stable and chaotic cases, a large stable region
at lower eccentricities, and a narrow stable island with moderate $e_b$, 
located about 1~$\sigma$ away from the global minimum. 

\subsubsection{Stable near-circular configuration}
\label{Stable near-circular configuration}

It is not surprising that the planetary system has better chances to survive with near circular orbits.
In these configurations, the bodies
might interact gravitationally, but at any epoch they will be distant enough to not exhibit close encounters.
By performing direct long-term N-body integrations, we conclude that 
the individual fits in the low-eccentricity region are stable for at least $10^8$ years,
and none of them is involved in low-order MMR. Instead, the average period ratio of 
the stable fits in this region is between 1.8 for the very circular fits and 1.88
at the border of the stable region.
This range of ratios is far away from the 2:1 MMR,
but could be close to a high-order MMR like 9:5, 11:6, 13:7, or even 15:8. However, we did not 
study these possible high-order resonances, and we assume that if not in 2:1 MMR, then the planetary system
is likely dominated by secular interactions.

The right column of Fig.~\ref{FigGam:1} shows the dynamical evolution of the most circular
fit from Fig.~\ref{FigGam:2} with
$\chi_{\mathrm{red}}^2$ = 14.1, ($\chi_{\mathrm{red}}^2$~=~1.22),
$m_{b} \sin{i_{b}}$ = 2.4 $M_{\mathrm{Jup}}$, $m_{c} \sin{i_{c}}$ = 3.2 $M_{\mathrm{Jup}}$,
$a_{b}$~=~1.28~AU, $a_{c}$~=~1.94~AU.
We started simulations with $e_{b,c}$~=~0.001, and the gravitational interactions forced the 
eccentricities to oscillate very rapidly between 0.00 and 0.06 for $\eta$~Cet~b, and between 0.00 to 0.03 for $\eta$~Cet~c. 
The arguments of periastron $\omega_b$~and~$\omega_c$ circulate between 0 and 360$^\circ$, but the secular 
resonant angle $\Delta\omega_{b,c}$~=~$\omega_{c}$~--~$\omega_{b}$, while circulating, seems 
to spend more time around 180$^\circ$ (anti-aligned).
Within the stable near-circular region, the gravitational perturbations between the planets have lower
amplitudes in the case of the most circular orbits than other stable fits with higher 
initial eccentricities (and~smaller~$\chi_{\mathrm{red}}^2$). 

The middle column of Fig.~\ref{FigGam:1} illustrates the best $\chi_{\mathrm{red}}^2$
fit within the low-eccentricity stable region with
$\chi_{\mathrm{red}}^2$~=~13.03 ($\chi_{\mathrm{red}}^2$~=~1.13)
and initial orbital parameters of $m_{b} \sin{i_{b}}$~=~2.4~$M_{\mathrm{Jup}}$, 
$m_{c} \sin{i_{c}}$ =~3.2~$M_{\mathrm{Jup}}$, 
$a_{b}$~=~1.28 AU, $a_{c}$~=~1.93 AU, $e_b$~=~0.06, and $e_c$~=~0.001.
The mean value of $\Delta\omega_{b,c}$ 
is again around 180$^\circ$, while the amplitude is $\approx\pm$~90$^\circ$. Immediately after the start of
the integrations $e_c$ has increased from close to 0.00 to 0.07, 
and oscillates in this range during the dynamical test, while $e_b$ 
oscillates from 0.05 to 0.11.
In particular, the highest value for $e_b$ is very interesting  
because, as can be seen from Fig.~\ref{FigGam:2}, starting integrations 
with 0.08 $< e_b <$ 0.11 in the initial epoch yields unstable solutions.
The numerical stability of the system appears to be strongly dependent
on the initial conditions that are passed to the integrator.
For different epochs the gravitational perturbations in the system would yield different orbital parameters than
derived from the fit, and starting the integrations from an epoch forward or backward in time where $e_b$ or $e_c$ 
are larger than the $e_{b,c}$ = 0.08 limit might be perfectly stable. 
 
We investigated the orbital evolution of a large number of fits in the low-eccentricity
region and did not find any aligned system configuration. 
Instead, all systems studied with near circular configurations seem to
settle in a secular resonance where $\Delta\omega_{b,c} \approx$ 180$^\circ$
shortly after the start of the integrations, and exhibit a semichaotic behavior.
This is expected as the system's secular resonance angle $\Delta\omega_{b,c}$ will circulate or librate,
depending on the initial $\Delta\omega_{b,c}$ and eccentricity values at the beginning of the stability test
\citep[e.g.,][]{Laughlin,Lee2003}.
The fits in the near circular island always favor $\Delta\omega_{b,c} \approx$ 180$^\circ$,
and thus the system spends more time during the orbital evolution in this anti-aligned configuration.

\subsubsection{Narrow 2:1 MMR region}
\label{Narrow 2:1 MMR region }

The low-eccentricity island is located farther away from the best co-planar fit than the 3$\sigma$ confidence level, so we 
can neither consider it with great confidence nor reject the possibility that the $\eta$~Cet system 
is perfectly stable in a near circular configuration.
Thus, we focused our search for stable configurations on regions closer to the best fit. 
In the edge-on coplanar $(e_b, e_c)$ grid (Fig.~\ref{FigGam:2}), we have spotted a few fits at the 1$\sigma$ border 
that passed the preliminary $10^5$ years stability test.
The additional long-term stability test proves that three out of four fits are stable for $10^8$ years.

To reveal such a set of stable orbital parameters, we created another high-density $(e_b, e_c)$ grid 
around these stable fits. We started with 0.131\,$\geq e_b \geq$\,0.181 
and 0.001\,$\geq e_c \geq$\,0.051 with steps of 0.001 (50x50 fits).
The significant increment of the resolution in this $(e_b, e_c)$ plane reveals a narrow stable island, where 
the vast majority of the fits have similar dynamical evolution and are stable for at least 
10$^8$ years (see left panel of Fig.~\ref{FigGam:2}).

The derived initial planetary periods for these fits are very close to those from the grid's best fit 
(\S\ref{Formally best edge-on coplanar fit}) and initially do not suggest any low-order MMR.
However, the mean planetary periods during the orbital evolution show that 
the system might be efficiently trapped in a 2:1 MMR. 
This result requires a close examination of the lowest order eccentricity-type resonant angles,

\begin{equation}
\theta_b =  \lambda_b - 2\lambda_c + \varpi_{b},~~~~\theta_c =  \lambda_b - 2\lambda_c + \varpi_{c}, 
\end{equation}

\noindent
where $\lambda_{b,c}$ =  $M_{b,c}$ + $\varpi_{b,c}$ is the mean longitude of the inner and outer planet, respectively.
The resonant angles $\theta_b$ and $\theta_c$ 
librate around $\sim$0$^\circ$ and $\sim$180$^\circ$, respectively, for the whole island,
leaving no doubt about the anti-aligned resonance nature of the system.
$\theta_b$ and $\theta_c$ librate in the whole stable region with very large amplitudes of nearly $\pm$180$^\circ$, so that
the system appears to be very close to circulating (close to the separatrix), but in an anti-aligned planetary 
configuration, where the secular resonance angle $\Delta\omega$ = $\theta_b$ - $\theta_c$ = $\omega_{b}$ - $\omega_{c}$ 
librates around 180$^\circ$.
A similar behavior was observed during the first 17\,000 years of orbital evolution of the best edge-on 
coplanar fit from \S\ref{Formally best edge-on coplanar fit}.
These results suggest that systems in the $(e_b,e_c)$ region around the best edge-on fit are close to 2:1 MMR, 
but also appear to be very fragile, and only certain orbital parameter combinations may lead to stability.

Fig.~\ref{FigGam:1} (left column) illustrates the orbital evolution of the system with the best stable fit from the resonant 
island; the initial orbital parameters can be found in Table~\ref{table:orb_par_stable}.
The eccentricities rapidly change with the same phase, reaching moderate levels of $e_b$ = 0.15\dots 0.25 and 
$e_c$ = 0.0\dots 0.08, while the planetary semimajor axes oscillate in opposite phase, 
between $a_b$ = 1.21\dots 1.29 AU and $a_c$ = 1.92\dots 2.06 AU.
During the dynamical test over 10$^8$ years $\Delta\omega_{b,c}$ librates around 180$^\circ$ with an 
amplitude of $\approx\pm$ 15$^\circ$,
while the $\omega_{b}$ and $\omega_{c}$ are circulating in an anti-aligned configuration.

\begin{figure}[htp]

\centering

\resizebox{\hsize}{!}{\includegraphics{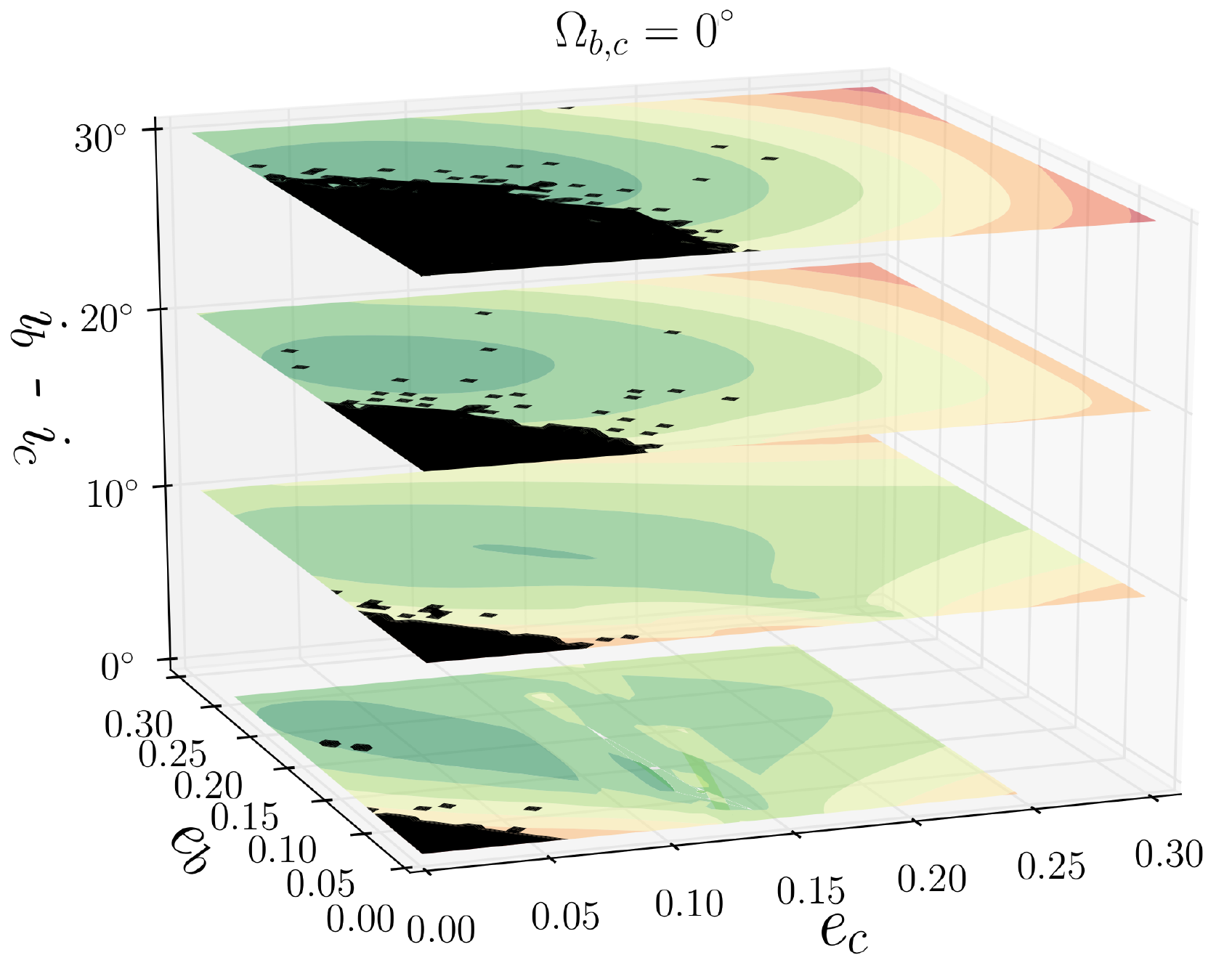}}

\caption{Mutually inclined grids where $\Omega_{b,c}$ = 0, and the mutual inclination comes only from
$\Delta i_{b,c}$ = $i_{b} - i_{c}$.
The low-eccentricity stable island increases its size
by assuming higher mutual inclinations, eventually creating an overlap with the 1$\sigma$ confidence level from the grid's best fit 
(dark green contours). 
 } 

\label{FigGam:MIT}%
\end{figure} 

\subsection{Coplanar inclined system}
\label{Coplanar inclined system}

To create coplanar inclined grids we used the same 
technique as in the edge-on coplanar grids, with additional constraints $i_{b}=i_{c}$ and 
$i_{b,c} \neq 90^\circ$.
We kept $\Omega_{b,c}$~=~0$^\circ$ as described in \S\ref{Formally best inclined fits},
so that we get an explicitly coplanar system.
A set of grids concentrated only on the 2:1 MMR stable island are shown in Fig.~\ref{FigGam:4}
(left panel), and larger grids 
for the low-eccentricity region are shown in the right panel of Fig.~\ref{FigGam:4}.
For reference, we placed the two grids from Fig.~\ref{FigGam:2} with $\sin{i}$ = 1 at the top of Fig.~\ref{FigGam:4}.
This coplanar inclined test shows how the stable islands behave if we 
increase the planetary masses via the LOS inclination.
We limited the grids with $\sin{i}\neq$ 1 in Fig.~\ref{FigGam:4} to a 
smaller region in the $(e_{b},e_{c})$ plane than the 
grids from Fig.~\ref{FigGam:2} to use CPU time efficiently, 
focusing on the potentially stable $(e_{b},e_{c})$ space
and avoiding highly unstable fit regions.

The smaller grid area is compensated for by higher resolution, however, as those grids have between 3\,600 and 10\,000 fits.
For simplicity we do not show the chaotic configurations, but illustrate only the 
stable fits that survived the maximum evolution span of 10$^5$ years (dark areas) for this test in Fig.~\ref{FigGam:4}.
We studied the near circular stable region by decreasing $\sin{i}$ with a step size of 0.05 in the range from 1 to 0.80.
The five stability maps of Fig.~\ref{FigGam:4} (right panel) clearly show the tendency 
that the near circular stable region becomes smaller when the planetary mass increases with decreasing $\sin{i}$. 
The stable island near circular orbits preserves its stability down to $\sin i \approx 0.75$ ($i\sim$~49 deg),
when even the most circular fit ($e_{b,c}$ = 0.001) becomes unstable.

The 2:1 MMR region strongly evolves and also decreases in size while decreasing $\sin{i}$. 
When $\sin{i}\approx$~0.94 ($i\approx70^\circ$), the 2:1 MMR region is smallest
and completely vanishes when $\sin{i}\leq$~0.93. 
We find that the largest stable area is reached for $\sin{i}$ = 1, and thus,
there is a high probability that the $\eta$~Cet system is observed nearly edge-on and involved in an anti-aligned 2:1 MMR.
The repeated tests with {\it SyMBA} for 2x10$^6$ years are consistent with the {\it Mercury} results,
although the stability regions were somewhat smaller.
This is due to the longer simulations, which eliminate the long-term unstable fits. Integrating for 10$^8$ years may leave only
the true stable central regions, 
however, such a long-term dynamical test over the grids requires much longer CPU time than we had for this study.

\subsection{Mutually inclined system}  
\label{Mutually inclined system} 

Constraining the mutual inclination from the RV data alone is very challenging, even for well-known and extensively 
studied extrasolar planetary systems, and requires a large set of highly precise RV and excellent astrometric data 
 \citep[see][]{Bean2}.
For $\eta$~Cet, the number of radial velocity measurements is relatively small, so that we cannot
derive any constraints on the mutual inclination from the RV. We tried to derive additional
constraints on the inclinations and/or the ascending nodes of the system from the Hipparcos Intermediate
Astrometric Data, as was done in \cite{Reffert2011} for other systems. We found that all but the lowest
inclinations (down to about 5~$^{\circ}$) are consistent with the Hipparcos data, so no further 
meaningful constraints could be derived.

We have shown in \S\ref{Two planet edge-on coplanar system} and \S\ref{Coplanar inclined system}  
that we most likely observe the $\eta$~Cet planetary system nearly edge-on, because we found a maximum of stable fits at $\sin{i}$ = 1,
in line with the Hipparcos constraint.
Assuming lower inclinations, the size of the stability region in the $(e_b,e_c)$ plane decreases. 
Because we constrained the system inclination only by stability criteria,
it would be interesting to see whether the stability will sufficiently increase if we allow a mutual inclination 
between the orbits to occur in our fits, or whether the system will become more unstable than for the coplanar fits.
Moreover, as we have shown in \S\ref{Formally best inclined fits}, by adding three additional fitting parameters
we obtain significantly better fits (although very unstable), 
and it is important to determine whether we can find any stable solutions  
or even stable islands for highly inclined non-coplanar configurations.

To study the dynamics of the $\eta$~Cet system for mutually inclined orbits we investigated the stability
with the following three different strategies:

1) We fixed the $\Delta i_{b,c}$ to be constant and assumed $\Omega_{b,c}$~=~0$^\circ$,
so that the mutual inclination depends only on $i_{b,c}$.
 We defined $\Delta i_{b,c}$~= $i_{b} - i_{c}$, where for the 2:1 MMR region the 
$i_{b}$ = 89.5$^\circ$, 89$^\circ$, 88.5$^\circ$ and
$i_{c}$ = 90.5$^\circ$, 91$^\circ$, 91.5$^\circ$. 
We found that the size of the 2:1 MMR stable region decreases very fast with mutual inclination, and
after $\Delta i_{b,c}>$ 2$^\circ$ the 2:1 MMR island completely vanishes.
For the global $(e_b, e_c)$ grid we defined $i_{b}$ = 85$^\circ$, 80$^\circ$, 75$^\circ$ and
$i_{c}$~=~95$^\circ$,  100$^\circ$, 105$^\circ$.  
The results from this test are illustrated in Fig.~\ref{FigGam:MIT}.
From the grid we found that the low-eccentricity stable region shows a trend of expanding its size
with the mutual inclination for $\Delta i_{b,c}$ = 10$^\circ$ and 20$^\circ$.
When $\Delta i_{b,c}$ = 30$^\circ$ 
the stable area expands and we find many stable fits for moderate $e_{c}$ within the 1$\sigma$ confidence region
from the $\Delta i_{b,c}$ = 30$^\circ$ grid. 
This test shows that for a high mutual inclination there is a high 
probability for the system to have near circular orbits, or moderate $e_{c}$.

\begin{figure}[htp]
\centering

\resizebox{\hsize}{!}{\includegraphics{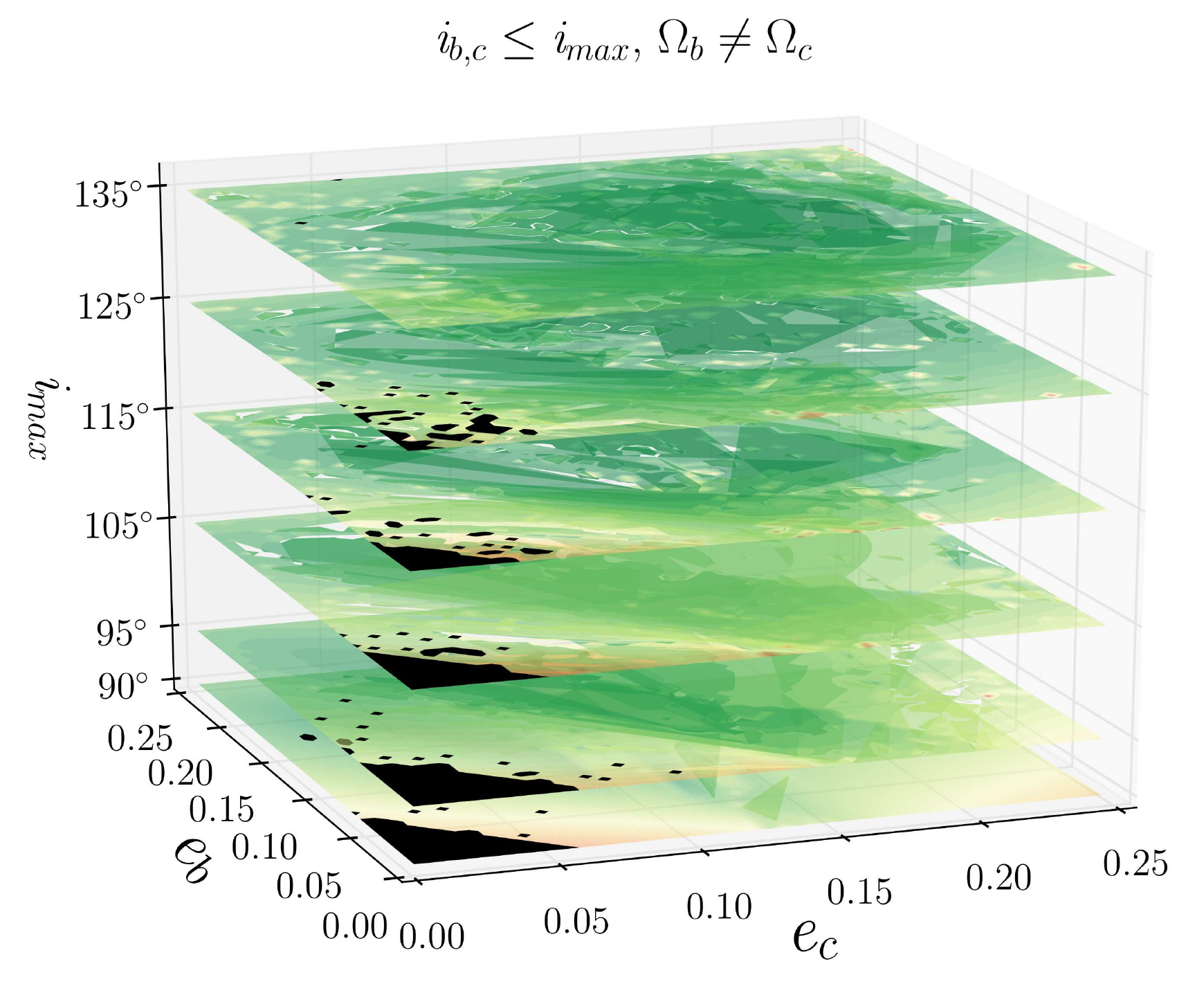}}
\caption{Grids with mutual inclination dependent on $\Delta\Omega_{b,c}$ and $i_{b,c} \leq i_{\mathrm{max}}$.
During the fitting the mutual inclinations from the grids are very low, on the order of $\sim 3^\circ$.
The stability area has a maximum at $i_{\mathrm{max}}$~=~95$^\circ$, slightly higher than the coplanar grid at 90$^\circ$.
The stability decreases at larger $i_{\mathrm{max}}$ due to the larger planetary masses and the stronger perturbations.
There is no stable solution at $i_{\mathrm{max}}$~=~135$^\circ$.
}

\label{FigGam:stable_inc}%
\end{figure}

2) We again relied on the $(e_b, e_c)$ grids, where additional free parameters in the fits 
are $i_{b,c}$ and $\Omega_{b,c}$. 
However, in the fitting we restricted the minimum inclination to be
 $i_{b,c}$ = 90$^\circ$, and the $\sin{i_{\mathrm{max}}}$ factor comes only from inclination
angles above this limit.
In this test we only studied the global $(e_b, e_c)$ plane, without examining the 2:1 MMR region separately.
Initially, for the first grid we set $\sin{i_{\mathrm{max}}}$ = 0.707 ($i_{\mathrm{max}}$ = 135$^\circ$), 
while $\Omega_{b,c}$ was unconstrained and was let to vary
across the full range from 0 to 2$\pi$. Later we constructed grids by
decreasing the maximum allowed inclination to $\sin{i_{\mathrm{max}}}$ = 0.819, 0.906, 0.966
and at last 0.996 ($i_{\mathrm{max}}$ = 125$^\circ$, 115$^\circ$, 105$^\circ$, and 95$^\circ$, respectively), 
thereby decreasing the upper limit on the planetary masses and mutual inclination angle $\Delta i_{b,c}$.
In all grids we allowed $i_{b} \neq i_{c}$, and $\Omega_{b}\neq\Omega_{c}$ as we discussed in 
\S\ref{Formally best inclined fits}. 
We find that when increasing $i_{\mathrm{max}}$, the grid's $\chi_{\mathrm{red}}^2$ values improved, and
the planetary $i_{b,c}$ are usually close to $i_{\mathrm{max}}$. The obtained $\Delta\Omega_{b,c}$ is also very low,
favoring low mutual inclinations in this test.
When $i_{\mathrm{max}}$ = 95$^\circ$ the average mutual inclination over the grid is 0.22$^\circ$, making the grid
nearly coplanar.
In this case the planetary masses are only $\sim$ 0.5\% higher than their minimum, which has a negligible effect on the stability. 
However, the very small orbital misalignment in those systems seems to have a positive influence on the system stability, 
and we find slightly more stable fits than for the edge-on 
coplanar case discussed in \S\ref{Two planet edge-on coplanar system} (see Fig.~\ref{FigGam:stable_inc}).

At larger $i_{\mathrm{max}}$ the $\chi_{\mathrm{red}}^2$ distribution in general has lower values, 
but the size of the stability region is decreasing.
This is probably due to the increased planetary masses and the stronger gravitational perturbations in the dynamical simulations.
There is not even one stable solution at $i_{\mathrm{max}}$~=~135$^\circ$, but we have to note that there are many $chaotic$ fits
that survived the 10$^5$ yr test, which are not shown in Fig.~\ref{FigGam:stable_inc}.

\begin{figure}[htp]
\centering

\resizebox{\hsize}{!}{\includegraphics{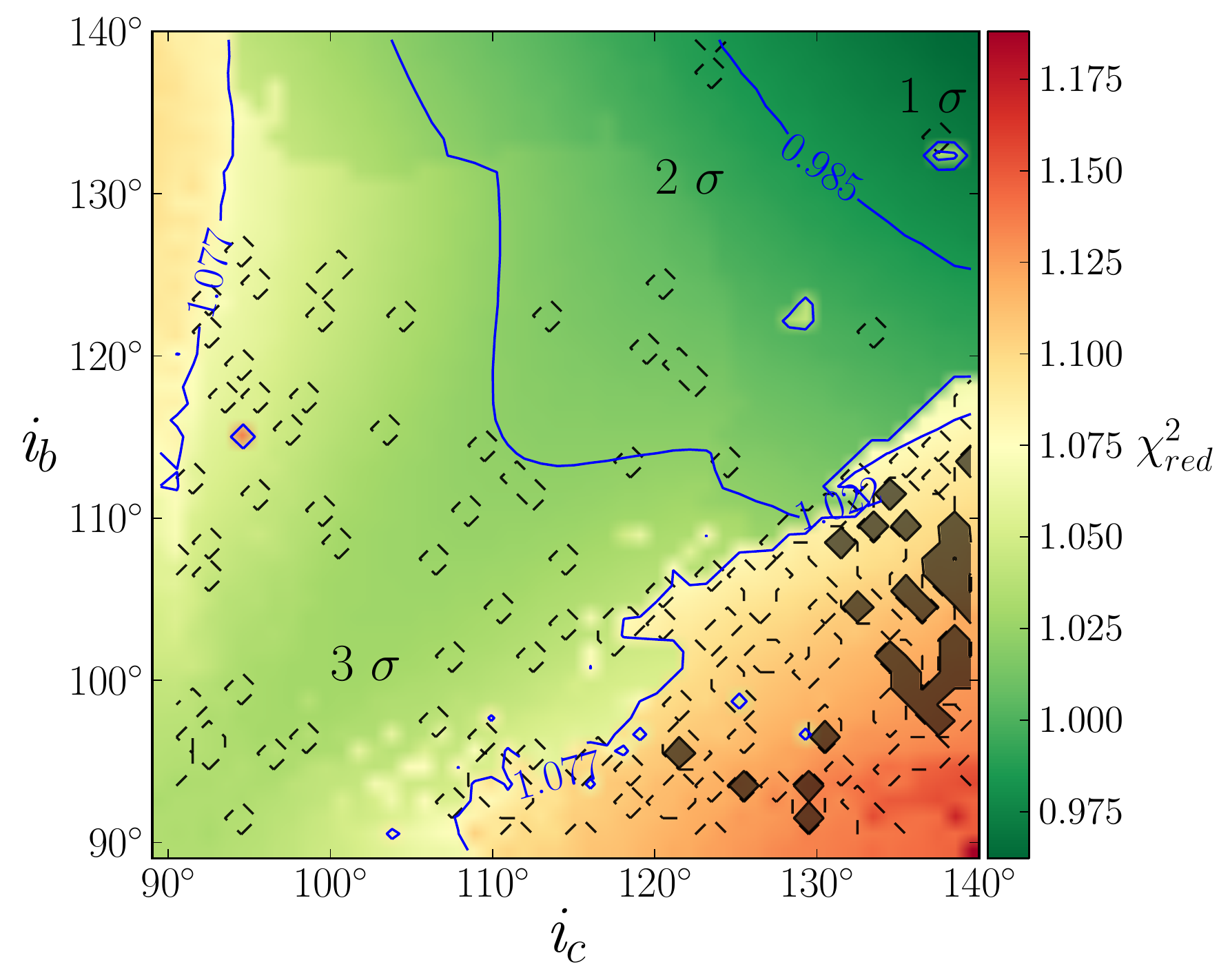}}
\caption{Except for the inclinations, the remaining orbital parameters are free to vary for 
the $(i_b, i_c)$ grid (50x50 fits). 
The $\chi_{\mathrm{red}}^2$ solutions suggest a higher LOS inclination and close to coplanar configurations. 
There are no obvious stable solutions near the coplanar diagonal of best fits from $i_{b,c}$ = 90$^\circ$ to 140$^\circ$).
Instead, there are many chaotic survivals and few stable islands 
at higher mutual inclinations, more than 3 $\sigma$ away from the grid minimum. }

\label{FigGam:i1_i2}%
\end{figure}

3) Because of the resulting low mutual inclination in the second test, we decided to 
decouple the planetary orbits and to test for higher mutual inclinations by allowing 
$\Delta\Omega_{b,c}$ $\neq$ 0, in contrast to the
first test where $\Delta\Omega_{b,c}$ = 0.
We constructed a grid of best fits for fixed $i_{b,c}$ ($i_b, i_c$ grid). 
The planetary inclination $i_{b,c}$ was increased from 90$^\circ$ to 140$^\circ$ with steps of 1$^\circ$, 
while rest of the orbital parameters were free in the fitting. 
This test attempts to check for stability for almost all the possible mutual inclinations in the system.
Fig.~\ref{FigGam:i1_i2} illustrates the stability output in the $(i_b, i_c)$ plane.
The $\chi_{\mathrm{red}}^2$ solutions in this grid have lower values when $i_b\approx i_c$ (close to the coplanar configuration),
and have a clear trend of better fits when the LOS inclination is high.
However, we could not find any stable configuration near the coplanar diagonal of best fits 
from $i_{b,c}$ = 90$^\circ$ to 140$^\circ$.
There are many chaotic survivals and few stable islands 
at higher mutual inclinations, more than 3$\sigma$ away from the grid minimum. 

We are aware of the fact that, when including the $i_{b,c}$ and $\Omega_{b,c}$ in the fitting model, the parameter space becomes
extremely large, and perhaps many more smaller and possibly stable minima might exist, 
but we have not been able to identify any in this study.

\subsection{Impact of stellar mass on the stability analysis} 
\label{Stellar mass impact on the stability analysis} 

Finally, we have tested how changing the assumed stellar mass influences the stability in $(e_b, e_c)$ space.
We generated coplanar edge-on $\chi_{\mathrm{red}}^2$ grids by assuming different stellar masses, 
using the values in Table.~\ref{table:phys_param}.
We took the same $(e_b, e_c)$ grid area and resolution as discussed in \S\ref{Two planet edge-on coplanar system} 
(Fig.~\ref{FigGam:2}) and started our grid search with 1~$M_\odot$ and 1.4~$M_\odot$, which are the lower and upper
stellar mass limits for  $\eta$~Cet proposed by \citet{Berio}.
Next we assumed 1.3~$M_\odot$ from \citet{Luck}, then 1.84~$M_\odot$, which is the upper limit 
from \citet{Reffert2014}, and we already have the 1.7~$M_\odot$ grid from \S\ref{Two planet edge-on coplanar system} .
The longest integration time applied to the stability test for these grids was 10$^5$ years.

\begin{figure}[htp]
\centering
\resizebox{\hsize}{!}{\includegraphics{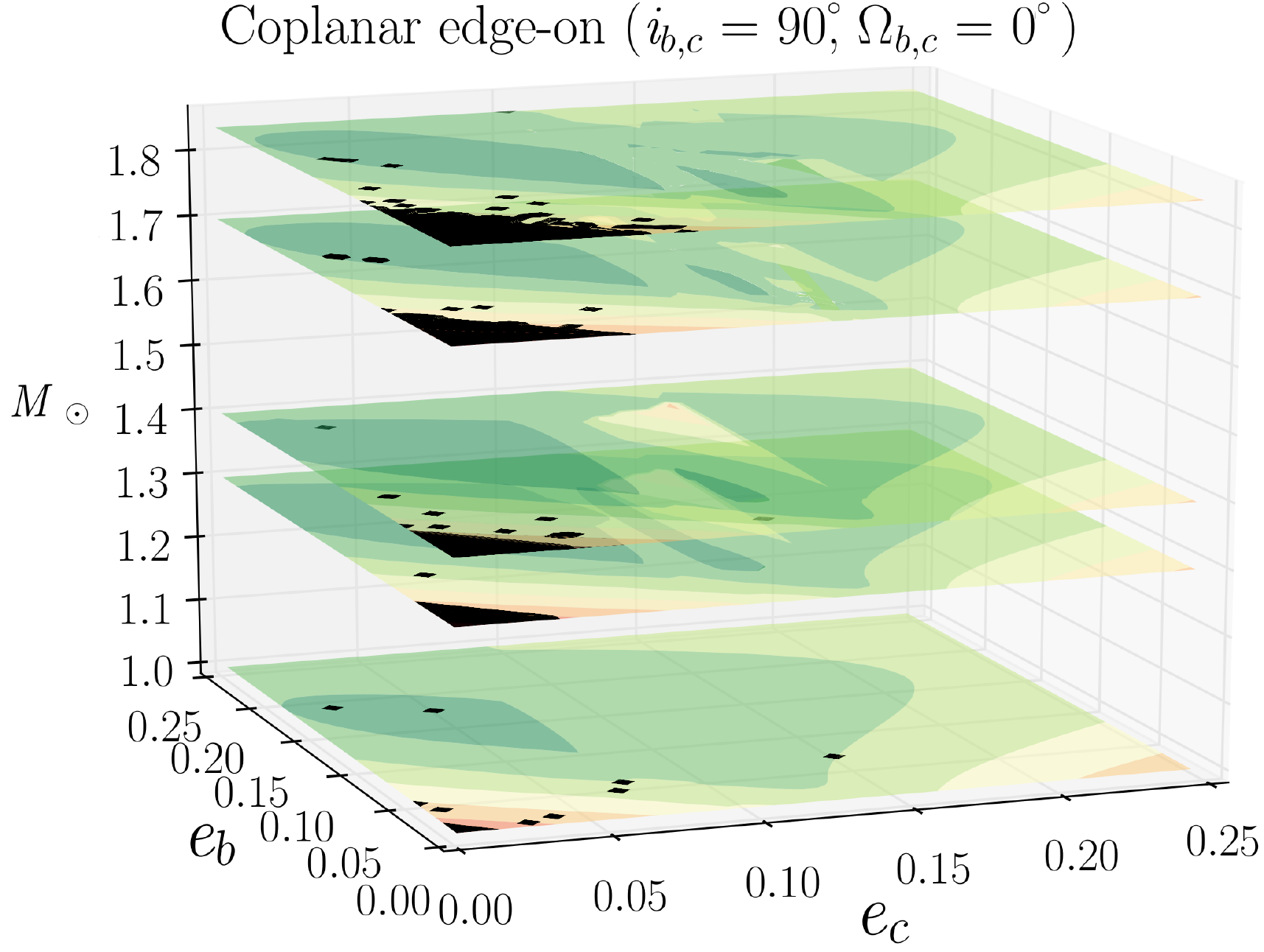}}

\caption{Stability maps with different initial masses for $\eta$~Cet.
A clear stability trend can be seen in the $(e_b, e_c)$ grids in the sense that by increasing the stellar mass
up to the maximum of 1.84 $M_\odot$, the size of the stable region increases as well.
} 
\label{FigGam:st_mass}%
\end{figure}

Stability results are shown in Fig.~\ref{FigGam:st_mass}, which clearly shows a trend of higher stability with larger stellar mass.
Our starting mass of 1~$M_\odot$ leads almost to the disappearance of 
the low-eccentricity stable island, and only a few stable solutions
can be seen for very circular orbits. Increasing the stellar mass to 1.3~$M_\odot$, 1.4~$M_\odot$ and 1.7~$M_\odot$
reveals larger stable areas at low eccentricities.
Finally, in the 1.84~$M_\odot$ grid we see the largest low-eccentricity stable island in our test.

Similarly, the 2:1 MMR region evolves and decreases in size with decreasing $\eta$~Cet mass. 
No resonant island is seen at 1~$M_\odot$.
However, the 2:1 MMR region has a larger area at 1.7 $M_\odot$ than in the 1.84 $M_\odot$ grid.

This effect comes from the fact that starting the fitting process with lower stellar mass
will also scale down the whole planetary system. 
The derived orbital angles and the periods from the dynamical fitting will remain almost the same in the 
$(e_b, e_c)$ grids when fitting from 1~$M_\odot$ to 1.84~$M_\odot$. However, as can be seen from 
equations (1), (3) and (4), the planetary masses, 
semimajor axes $a_{b,c}$, and the Hill radii $r_{b,c}$ are dependent on the stellar mass.
By scaling down the planetary masses, we would expect the gravitational interactions between the planets to be less
destructive, and we would thus expect more stable fits when adopting a lower mass for the primary. 
This is exactly the opposite of what can be seen in Fig.~\ref{FigGam:st_mass}. 
The dynamical simulations are much more sensitive to $\Delta a_{b,c}$ = $a_c - a_b$
than to the planetary mass ratio.
From equations (4) and (5) it can be seen that $r_{b,c}$ will slightly decrease when
assuming a lower stellar mass, although not as fast as $a_{b,c}$ will decrease.
$\Delta a_{b,c}$ for the 1~$M_\odot$ grid is on average $\approx$ 0.54 AU, 
and for the 1.84~$M_\odot$ grid $\approx$ 0.65 AU. 
The more similar planetary semimajor axes in lower stellar mass systems lead to a higher number of
close encounters during the orbital evolution,
and thus, to a higher ejection rate, especially for fits with higher $e_{b,c}$. 
On the other hand, the $\Delta a_{b,c}$ for the grid with maximum mass for $\eta$~Cet (1.84~$M_\odot$)
is enough to keep the two planets well separated,
and the stability region in lower eccentricities increases.

\section{Discussion}
\label{Discussion eta cet}

\subsection{Planetary hypothesis} 

In principle, the possible reasons for observed RV variability in K giants 
are rotational modulation of star spots, long-period nonradial pulsations, 
or the presence of planets.

Star spots can most likely be excluded as a viable explanation for the observed RV
variability of $\eta$~Cet, at least for one of the two observed periods.
If star spots were to cause the RV to vary,
one of the two periods that we observe would have to match the 
rotational period (still leaving the second period unexplained).
However, the longest rotation period of $\eta$~Cet compatible
with its radius of $R$~=~14.3~$\pm$~0.2~$R_{\odot}$ and its projected rotational velocity
$v \sin i$~=~3.8~$\pm$~0.6~km\,s$^{-1}$ \citep{Hekker2} is 190$_{-26}^{+36}$~days. 
This is much shorter than either the 405-day or the 750-day period. 
Moreover, one would expect a larger photometric variation than the 3~mmag 
observed by Hipparcos for a star spot to generate an RV amplitude on the order
of 50~m\,s$^{-1}$ \citep{Hatzes2000}.\footnote{The Hipparcos observations were taken earlier 
(until 1993) than 
our RV data (since 2000), but assuming that $\eta$~Cet was quiet for four years, and active with a very
constant RV amplitude for more than ten years just seven years later appears rather contrived.}
On the other hand, macro-turbulent surface structures on 
K giants are currently poorly understood. Large and stable convection cells could act as
velocity spots, yielding an RV variability without significant photometric variability 
\citep[e.g.,][]{Hatzes2000}. Although unlikely, we cannot fully exclude that
the shorter 405-day period is due to rotational modulation of surface features, while the 
longer 750-day period is most certainly due to a planet.

Ruling out nonradial pulsations is harder than ruling out star spots, but 
also possible. First, we see evidence of eccentricity in the shape
of the radial velocity curves, which is an indication for a Keplerian
orbit. Second, the signal has been consistent over 12 years; this is not
necessarily expected for pulsations. Third, on top of the optical RV data we
derived the RV from the infrared wavelength regime (CRIRES). 
Although the IR RV error is much larger than that from the optical data, the CRIRES data are
clearly consistent with the Lick data. 
The best dynamical fit derived from the visible data and applied
only to the near-IR data 
has $\chi^2_{\rm red}$~=~0.502 and $r.m.s.$~=~26.7~m\,s$^{-1}$, while 
a constant model assuming no planets has $\chi^2_{\rm red}$ = 2.806  and $r.m.s.$~=~63.2~m\,s$^{-1}$. 
Therefore, the two-planet fit is more likely.
Just based on the CRIRES data, we
can rule out infrared amplitudes smaller than 30~m\,s$^{-1}$  or larger than 65~m\,s$^{-1}$
with 68.3\% confidence. If the RV variability in the optical were to be caused 
by pulsations, one might expected to find a different amplitude in the IR, 
which is not the case.
Fourth, there are no indications for these long-period nonradial pulsations in
K giants yet, although a variety of very short pulsation modes have been found
with Kepler \citep{Bedding2011}. Taken together, there is no supporting evidence
for the presence of pulsations in $\eta$ Cet.

The strongest evidence supporting a multiple planetary system comes from the dynamical 
modeling of the RV data.  
Taking the gravitational interactions between the two planets into account leads to a 
considerably better fit than the two Keplerian model (or two sinusoids). In other words, 
we were able to detect the interactions between the two planets in our data, 
which strongly supports the existence of the two planets.

Thus, though we cannot completely rule out alternative explanations, a two-planet
system is the most plausible interpretation for the observed RV variations of $\eta$ Cet.

\subsection{Giant star planetary population}  

With the detection of the first extrasolar planet around the giant star $\iota$~Dra~b 
(the first planet announced from our Lick sample -- \citet{Frink2})
the search for planets around evolved intermediate mass stars has increased very rapidly.
Several extrasolar planet search groups are working in this field to provide important
statistics for planet occurrence rates as a function
of stellar mass, evolutionary status, and metallicity
\citep[][\& etc.]{Frink,Sato2,Hatzes,Niedzielski2,Doellinger,Mitchell}. 
Up to date, there are 56 known planets around 53 giants stars in the literature, and all of them are in wide orbits.

Except for the $\eta$~Cet discovery, there are only three candidate multiple planetary systems known to orbit giant stars,
and all of them show evidence of two massive substellar companions.
\citet{Niedzielski_a,Niedzielski_b} reported 
planetary systems around the K giants HD~102272 and BD~+20~2457.
While the published radial velocity measurements of HD~102272 are best modeled with a double Keplerian,
 the sparse data sampling is insufficient to derive an acurate orbit for the second planet.
Initial stability tests based on the published orbital parameters show a very fast collision between the planets.
This appears to be due to the presumably very eccentric orbit of the outer planet, which leads to close encounters,
making the system unstable.
The case of BD~+20~2457 is even more dramatic. The best-fit solution suggests a brown dwarf and a very 
massive planet just at the brown dwarf mass border,
with minimum masses of $m_{1}\sin{i_{1}}\approx$ 21.4 $M_{\mathrm{Jup}}$
and $m_{2}\sin{i_{2}}\approx$ 12.5 $M_{\mathrm{Jup}}$, 
respectively (1 is the inner and 2 the outer planet).
This system has very similar orbital periods of $P_{1}\approx$ 380 days and $P_{2}\approx$ 622 days for 
such a massive pair, and the gravitational interactions are extremely destructive.
We have tested several orbital parameter sets
within the derived parameter errors and, without conducting any comprehensive stability analysis, so far
we were unable to find any stable configuration for BD~+20~2457. Recently, \citet{Horner2014} 
investigated the stability of BD~+20~2457 in more detail and did not find any stable solutions either.
However, we have to keep in mind that the formal best fit for $\eta$~Cet was also unstable, and only 
after an extensive stability analysis did we find long-term stable regions.
In this context we do not exclude the possibility that HD~102272 and BD~+20~2457 
 indeed harbor multiple planetary systems, but better data sampling is required 
to better constrain the planetary orbits. Moreover, additional efforts must be undertaken 
to prove the stability for these two systems, perhaps including highly mutually inclined 
configurations, or even better constraints on the stellar masses.

Of particular interest is also the system around the K~giant $\nu$~Oph \citep{Quirrenbach,Sato}, which is 
consistent with two brown dwarfs with masses $m_{1}\sin{i_{1}}\approx$ 22.3~$M_{\mathrm{Jup}}$, 
$m_{2}\sin{i_{2}} \approx$ 24.5~$M_{\mathrm{Jup}}$.
The two brown dwarfs exhibit a clear 6:1~MMR, with periods of $P_{1}\approx$ 530 days
and $P_{2}\approx$ 3190 days.

Although $\nu$~Oph and potentially BD~+20~2457 are not planetary systems, they
may present important evidence for brown dwarf formation in a circumstellar disk.
Such objects may form because in general the more massive stars should have more massive disks from which 
protoplanetary objects can gain enough mass to become brown dwarfs.
It might be possible for the 6:1 resonance configuration of $\nu$~Oph to have formed
via migration capture in a protoplanetary disk around a
young intermediate-mass progenitor, and the brown dwarf occurrence may be rather high
 \citep{Quirrenbach}.

Therefore, if we exclude $\nu$~Oph, which is clearly a brown dwarf system,
and the HD~102272 and BD~+20~2457 systems, which suffer from poor data sampling and stability problems, 
$\eta$~Cet is currently the only K giant star that shows strong evidence for harboring a 
stable multiplanetary system.

\subsection{Unique orbital configuration of $\eta$ Cet} 

The stable solutions from the 2:1 MMR region in the edge-on coplanar 
and inclined tests raise some important questions about the possible 
formation and evolution of the $\eta$~Cet planetary system. 
The $\theta_{b} \approx 0^\circ$ and $\theta_{c} \approx 180^\circ$ 2:1 MMR 
configuration is similar to that of the 2:1 resonance between the Jovian satellites Io and Europa,
but the Io-Europa configuration is not supposed to exist for relatively large eccentricities 
\citep[see][]{Beauge2003,Lee_m}. 
The average $e_{b}$ ($\sim 0.2$) and $e_{c}$ ($\sim 0.05$) for 
$\eta$~Cet are much larger than the $e_{b}$,$e_{c}$ boundary 
where the $\theta_{b} \approx 0^\circ$ and $\theta_{c} \approx 180^\circ$ configuration exists. 
An aligned configuration with both $\theta_{b}$ and $\theta_{c}$ librating about $0^\circ$ is expected for a mass ratio
$m_{b}/m_{c} \approx 0.77$ and the eccentricities in the 2:1 MMR region.

One possible stabilizing mechanism for the $\eta$~Cet system might be the large libration 
amplitudes of both $\theta_{b}$ and $\theta_{c}$, which are almost circulating. 
We made some preliminary attempts and failed to find a small libration amplitude counterpart.
More work is needed to understand the stability of this 2:1 MMR configuration, as well as its origin, 
if $\eta$~Cet is indeed in this configuration.

We cannot fully exclude that the true system configuration of $\eta$~Cet corresponds to some of the 
single isolated stable fits that we see in the stability maps, and neither can we exclude one of 
the numerous fits that are stable for 10$^8$ years, which show a scattering chaotic behavior.
A nonresonant system in near circular orbits is also possible.

\section{Summary}
\label{Summary}

We have reported the discovery of a planetary system around the K giant star $\eta$~Cet. 
This discovery is the result of a long-term survey, which aims to discover planetary 
companions around 373 intermediate-mass G and K giant stars, and which started back in 1999 at Lick Observatory.
We presented 118 high-precision optical radial velocities based on the observations
with the Hamilton spectrograph at Lick Observatory 
and nine near-infrared data points from the ESO CRIRES spectrograph; these data cover more than a decade. 

We have fitted a dynamical model to the optical data, which ensured that any possible gravitational interactions 
between the planets are taken into account in the fitting process.
We showed that the dynamical model represents a significant $\chi_{\mathrm{red}}^2$ 
improvement over the double-Keplerian fit. 

In an attempt to characterize the most likely planetary configuration,
we performed an extensive stability analysis of the $\eta$~Cet system. 
We made a wide variety of high-resolution co-planar and inclined dynamical $\chi_{\mathrm{red}}^2$ grids,
which we used as an input for our numerical analysis. 
Thus, we transformed these grids into detailed stability maps.
In total, we carried out more than 200\,000 dynamical integrations with 
typical time spans of $10^5$ and 2x10$^6$ years, and we 
extended the test to $10^8$ years to study the edge-on coplanar case.

For the edge-on coplanar grid we used a set of best fits for fixed $e_{b}$ and $e_{c}$. 
We found that the $\eta$~Cet system can be stable for at least $10^8$ years, locked 
in a 2:1 MMR in a region with moderate $e_b$, which lies about 1$\sigma$ away from the best co-planar fit.
A much larger nonresonant stable region exists with nearly circular orbits, 
although it is located more than 3$\sigma$ away from the best fit and is thus less likely.
In the 2:1 MMR region all fits are in an anti-aligned planetary configuration and very close to the 
separatrix. The low-order eccentricity-type resonant angles $\theta_b$ and $\theta_c$ librate around 0$^\circ$
and 180$^\circ$, respectively, but with very large amplitudes of $\approx\pm$ 180$^\circ$.  
A similar near-separatrix behavior can be seen in the stable fits with near circular orbits, 
where the secular resonance angle $\Delta\omega_{b,c}$ circulates, but during most of the simulation
the planetary configuration is anti-aligned ($\Delta\omega_{b,c}\approx$ 180$^\circ$).

We provided a detailed set of $(e_{b}, e_{c})$ coplanar inclined stability maps, showing that the $\eta$~Cet
system is very likely observed in a near edge-on configuration ($i_{b,c}\approx$ 90$^\circ$).
The size of the stable region is largest when the system is assumed to have $i_{b,c}$ = 90$^\circ$, and 
when we increased the planetary mass via $\sin{i_{b,c}}$,
the size of the two stability regions in the $(e_{b}, e_{c})$ plane decreased.
The 2:1 MMR stable island totally disappeared when $i_{b,c}\approx$ 70$^\circ$, while the near circular stable island survived until
the LOS inclination becomes $i_{b,c}\sim$~49$^\circ$. Below the last inclination limit, all the fits 
in the $(e_{b}, e_{c})$ plane became unstable. 

We also presented results from a grid based on mutually inclined configurations, 
although we pointed out that they need not be exhaustive.
This is because of the large amount of computational time needed when the parameter space is expanded.
Another way of constraining the mutual inclination would be a very extensive and precise radial 
velocity and astrometric data set, which is so far not available for $\eta$~Cet. 

The most important conclusion from the inclined dynamical test is that the planets 
cannot be more massive than a factor of
$\sim$1.4 heavier than their suggested minimum masses. 
Higher inclinations, and thus larger planetary masses, lead to instability in all cases.
This excludes the possibility of two brown dwarfs purely based on stability considerations,
and strongly favors a planetary system.

We also tested how the uncertainty of the stellar mass will affect the dynamical stability of the system.
Decreasing the stellar mass leads to a smaller size of the stable region in the $(e_{b}, e_{c})$ grids, 
and thus we conclude that the stellar mass value from \citet{Reffert2014} is indeed a very reasonable estimate.

The $\eta$~Cet system is only the fourth candidate multiple substellar system around a G or K giant star 
and presents an important milestone for understanding planetary formation and evolution as a 
function of stellar mass, metallicity, and age. 

\begin{acknowledgements}
Part of this work was supported by the International Max Planck Research School for
Astronomy and Cosmic Physics at the University of Heidelberg,
IMPRS-HD, Germany. M.H.L. was supported in part by the Hong Kong RGC grant HKU 7024/13P.

We would like to thank the staff at Lick Observatory for their support over the years of this project. 
We kindly thank the CAT observers that assisted with this project, including Saskia Hekker,
Simon Albrecht, David Bauer, Christoph Bergmann, Stanley Browne, Kelsey Clubb, Dennis K\"{u}gler,
Christian Schwab, Julian St\"{u}rmer, Kirsten Vincke, and Dominika Wylezalek. 
We thank Mathias Zechmeister and Ansgar Reiners for their help with the acquisition and reduction of the CRIRES data. 
We thank Stefano Meschiari for the very helpful discussion regarding the capabilities of the Console package.
We would also like to thank our referee, Artie Hatzes, for his constructive comments 
that helped to improve this paper.     
\end{acknowledgements}

\bibliographystyle{aa}
\bibliography{eta_cet}

\end{document}